# An Integrated SERVQUAL–Lean Six Sigma Framework for Measuring Customer Satisfaction in Computer Service Companies

by

Mohammed Abboodi
BS. University of Baghdad, 2008

A thesis submitted in partial fulfillment of the requirements
for the degree of Master of Science
in the Department of Industrial Engineering and Management Systems
in the College of Engineering and Computer Science
at the University of Central Florida
Orlando, Florida

Spring Term
2014

Major Professor: Ahmad K. Elshennawy

# ABSTRACT

The computer service industry has been expanding dramatically due to the increase in the number of computing machineries in the last two decades. The entrance of large size companies in the market and the release of online tools that have the ability for diagnosing and troubleshooting hardware and software issues have boosted the competition. In the meantime, many of the small and medium size companies find themselves unable to keep their customers satisfied since their competitors provide high quality service with lower cost.

The lack of a good measurement system to assess and analyze the satisfaction level with the provided service is the fundamental cause of customer decline. The aim of this study is to construct a robust framework to measure customer satisfaction and highlight the root causes of dissatisfaction in the computer service sector. This framework brings together the key aspects of Six Sigma and SERVQUAL instruments into a structured approach to measure and analyze customer satisfaction with computer services. It deploys the DMAIC problem solving methodology along with the SERVQUAL model, which contributes service dimensions and the Gap Analyze technique.

Literature review indicates there have not been enough studies conducted to integrate Lean Six Sigma with SERVQUAL. To explore the effectiveness of the current framework, a computer service company has been selected. The satisfaction levels are calculated and the root causes of dissatisfaction have been identified. With a low overall customer satisfaction level, the company did not fulfill their customer requirements due to five major causes. Eliminating those causes will boost customer satisfaction, reduce the cost of acquiring new customers and improve the company performance in general.

# ACKNOWLEDGMENT

To begin with, I appreciate the guidance and support of the Department of Industrial Engineering and Management Systems at the University of Central Florida. My appreciation also extends to the faculty and staff who aided me in completing my Master's Thesis endeavor.

I have a deep gratitude to my advisor Dr. Ahmad Elshennawy for his patience, guidance and encouragement. I also want to thank him for the time he spent advising me as to my research which became a polished, refined study to make it valuable. My gratitude is extended as well to the committee members Dr. Luis Rabelo and Dr. Petros Xanthopoulos for their support and participation in the development of this study.

# CHAPTER 1
# INTRODUCTION

Today, customer satisfaction has become the key for company success and survival in global competition. Yet, many studies on customer behavior show that pleasing existing customers is easier than attracting new customers. Satisfying the customer is not an easy process and many studies show it is multidimensional whereas the growing of online transactions and online computer services has a huge impact on customer expectations and loyalty. In other words, the customers become smarter and more demanding.

Recently, many online tools (websites) have been developed to compare products or services with competitors' – many companies use these tools in their websites as advertisements to show that their products or services are either cheaper or better than their competitors. Most computer service providers such as Asseco Poland SA, Click Computers and BestBuy use online customer reviews and display those reviews online as marketing tools to attract new customers. The retailer can use this feedback to improve their performance and increase the customer satisfaction. Since many of the unsatisfied customers will provide feedback to future customers, the retailer should ensure the majority of their customers are satisfied before publishing the feedback. Otherwise, this may build an atrocious reputation for the company.

Online tools have also been developed for diagnosing and repairing computers, networks or even maintaining databases. Many larger organizations such as HP, Dell and Sony have opened the competition to smaller companies by offering online and on-site services. Competition in the computer service industry has increased and the chance for survival of small businesses has been reduced. To beat those big players, the firm should excel in its service quality. The company must be innovative to survive and sustain the competitive advantage in global competition. To be innovative they must use a strategic plan to identify customer needs and requirements, and satisfy them later. However, it is not easy to measure and satisfy those requirements since most of the service characteristics are intangibles.

Lean Six Sigma is a powerful tool deployed in many industrial applications such as manufacturing and engineering management. It leads to massive savings and performance improvement for many organizations. The author believes applying this technology will lead to a significant enhancement in customer satisfaction. Lean Six Sigma utilizes many tools and techniques such as quality function deployment (QFD), value stream mapping, continuous improvement and various quality techniques used to identify customer requirements and needs. By applying Lean Six Sigma, the customer satisfaction will dramatically increase. By identifying and analyzing the voice of customers, the designer and manufacturer will gain insight into the pros and cons of the product or service.

To sum up, this study investigates the dimensions of customer satisfaction and contributes a framework to improve the quality of the service in the computer service industry via pinpointing the root cause of customer dissatisfaction. By deploying this framework, the organization will boost their profit through increasing the number of satisfied customers along with reducing the relevant cost of marketing expenses.

# CHAPTER 2
# LITERATURE REVIEW

This chapter has been developed to explore the current literature that has been made in the area of this research. At the beginning, the author provides an overview for publications that discussed customer satisfaction. Also many customer satisfaction measurement tools have been explored. Then, the author has moved to describe the service quality and the models that have been used to measure it. Since the objective of the study is to optimize the customer satisfaction in computer service industry, a large part of this chapter has been specified to discuss the SERVQUAL and Gap analysis. Also, Lean Six Sigma and DMAIC have been described in detail.

## 2.1. Customer satisfaction:

Today, satisfying the customer is the highest priority of many global companies, whereas recent literatures indicate that customer satisfaction is a critical factor to achieve company long term success. In other words, it makes the organizations more competitive and more successful. Furthermore, many experts show that satisfying and keeping current customers are far less expensive than constantly replacing those (Jonson & Gustasson, 2000). Satisfying the customer not only saves money and increases profit but also brings repetitive and new business. Based on Churchill and Surprenant, the concept of customer satisfaction becomes a core of marketing research and practices (Churchill & Surprenant, 1982). Customer satisfaction can be defined as, "judgment that product or service feature, or service itself, provided a pleasurable level of consumption–related fulfillment, including levels of under or over fulfillment" (Oliver, 1997), or as "emotional response associated with particular product or services purchase, retail, outlet or molar pattern of behavior as well as market place" (Westbrook and Reilly, 1983). Many literatures show the main reason for customer decay is the product or service provider failed to identify and satisfy the customer needs accurately. To achieve high customer satisfaction, the organizations must implement well-designed questionnaire programs. They must also implement a professional process to collect and analyze the data.

Many researchers argue that customer satisfaction has a direct impact on customer loyalty (Rust & Zahorik, 1993). Furthermore, Shankar, Smith and Rangaswamy confirm that customer satisfaction and customer loyalty are positively correlated (Shankar, Smith and Rangaswamy, 2003). Shoemaker and Lewis argue that customer satisfaction is not equal to customer loyalty. Allen and Tanniru suggest that customer satisfaction is necessary to achieve customer loyalty, but it doesn't guarantee it (Allen and Tanniru, 2000). For instance, research that has been conducted by the University of Michigan to study the relationship between satisfaction and loyalty in the automobile industry shows Cadillac received first place in customer satisfaction. However, Cadillac's customers do not have a strong commitment toward the product which shows not all satisfied customers are loyal.

Many publications aim to provide a comprehensive definition for loyalty. Looy, Gemmel & Dierdonck define customer loyalty as, "customer behavior characterized by a positive buying pattern during an extended period (measured by means of repeat purchase, frequency of purchase, wallet share or other indicators) and driven by a positive attitude towards the company and its products or services" (Looy, Gemmel & Dierdonck, 2003). Oliver proposes yet another definition for customer loyalty, describing it as, "a deeply held commitment to rebuy or re-patronize a preferred product/service consistently in the future, thereby causing repetitive same-brand or same brand-set purchasing, despite situational influences and marketing efforts having the potential to cause switching behavior" (Oliver, 1999). A lot of literature emphasizes customer loyalty since it has an important role for organizations' success. Furthermore, a number of studies show there is a strong relationship between customer loyalty and organizational profit. In 1990, Reichheld and Sasser stated that customer loyalty has either positive or negative impact on the numbers of purchases, the cost of acquiring new customers and number of repeated purchases (Reichheld & Sasser, 1990). For instance, loyal customers buy more and that frequently reduces the cost of acquiring new customers. In addition, a study was conducted by the same authors and it showed that organizations may increase their profit 25% - 85% by retaining just 5% more of its customers. There are many types of customer loyalty. Based on Hill and Alexander, customer loyalty can be categorized in five varieties: Monopoly loyalty, cost of change loyalty, incentivized loyalty, habitual loyalty and committed loyalty. Moreover, many authors argued that customer loyalty has two dimensions (Engel & Blackwell, 1980; Julander, 1997). The first dimension is called behavioral dimension, which reflects the customers' repeated behavior to purchase the product (Kandampully & Suhatanto, 2000). Attitudinal dimension is the second dimension, which exhibits customer commitment to repurchase and promote the brand.

### 2.1.1 Measure customer satisfaction, loyalty:

Without question, measuring customer satisfaction is a critical step not only to achieve customer satisfaction but also to enhance customer support. Whereas, measuring customer satisfaction will direct the organizations to generate effective ways to improve their product qualities (Perkins, 1993). Smith identified four fundamental customer satisfaction measurements labeled: Overall Satisfaction Measurements, Loyalty Measurements, a Series of Attribute Satisfaction Measurements and Intentions to Repurchase Measurements (Smith, 2012).

Kristensen, Kanji and Dahlgaard came up with a framework to measure customer satisfaction. Their framework can be summarized in seven steps. The first step is identifying the product quality characteristics that increase customer satisfaction. The customer population should be defined in the second step. Later on, the decision should be made to identify whether the company is going to sample the total potential market or just the existing customers. Then the framework for the sampling should be constructed. Later on, the questionnaires should be designed and appropriate scales are made to measure customer satisfaction. Next, the surveys should be conducted which can be done by personal interviews, telephone interviews, or electronic surveys. Statistical tools are used to organize and analyze the collected data from the previous step after the results are communicated. This framework has been deployed in many European organizations work very efficiently (Kristensen, Kanji & Dahlgaard, 1992).

However, Perkins argued that measuring customer satisfaction could be done in three steps. Product dimensions on which each product varies should be identified first. Then, ask the customer to rate those dimension comparisons to other companies. At the end, the customer should be asked the overall satisfaction rating of the company (Perkins, 1993).

Hays developed a general model for development and use of customer satisfaction which consists of three steps as shown in Figure 2.1. The first phase is designated to define key characteristics of the products or service. At this step, the organization will get a better understanding of what customers need. The second step is specified to develop a questionnaire to assess customer's perception. Using the developed questionnaire is the last step (Hays, 1997).

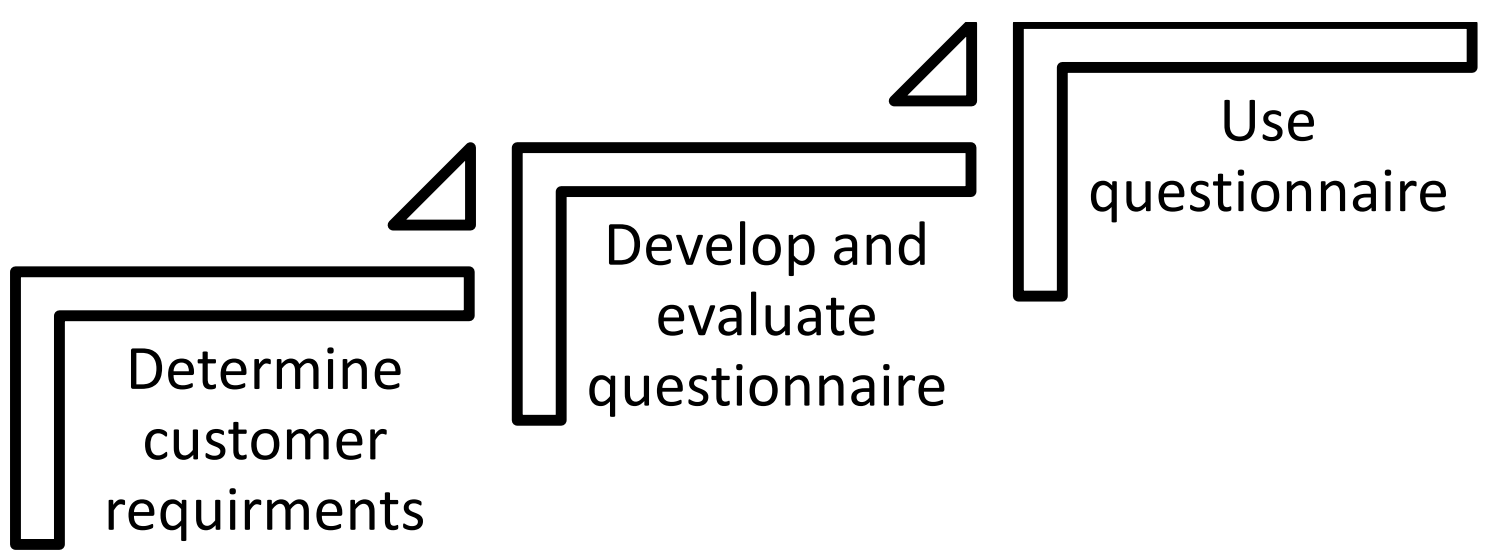

Figure 2.1: Hays Model to Measure Customer Satisfaction

Another model to measure satisfaction level is Customer Satisfaction Index (CSI) which was established in 1989 for measuring and evaluating the organization performance satisfying the customers' requirements and needs. However, each country has its own indicator to measure customer satisfaction. In 1994, the American Customer Satisfaction Index was introduced – 200 organizations from 34 different industries were surveyed and the results were reported (Fornell, 1996). The ACSI explains customer satisfaction as the outcome of three key elements: perceived quality, expectation and perceived value (Vavra, 1997). Figure 2.2 illustrates the ACSI model.

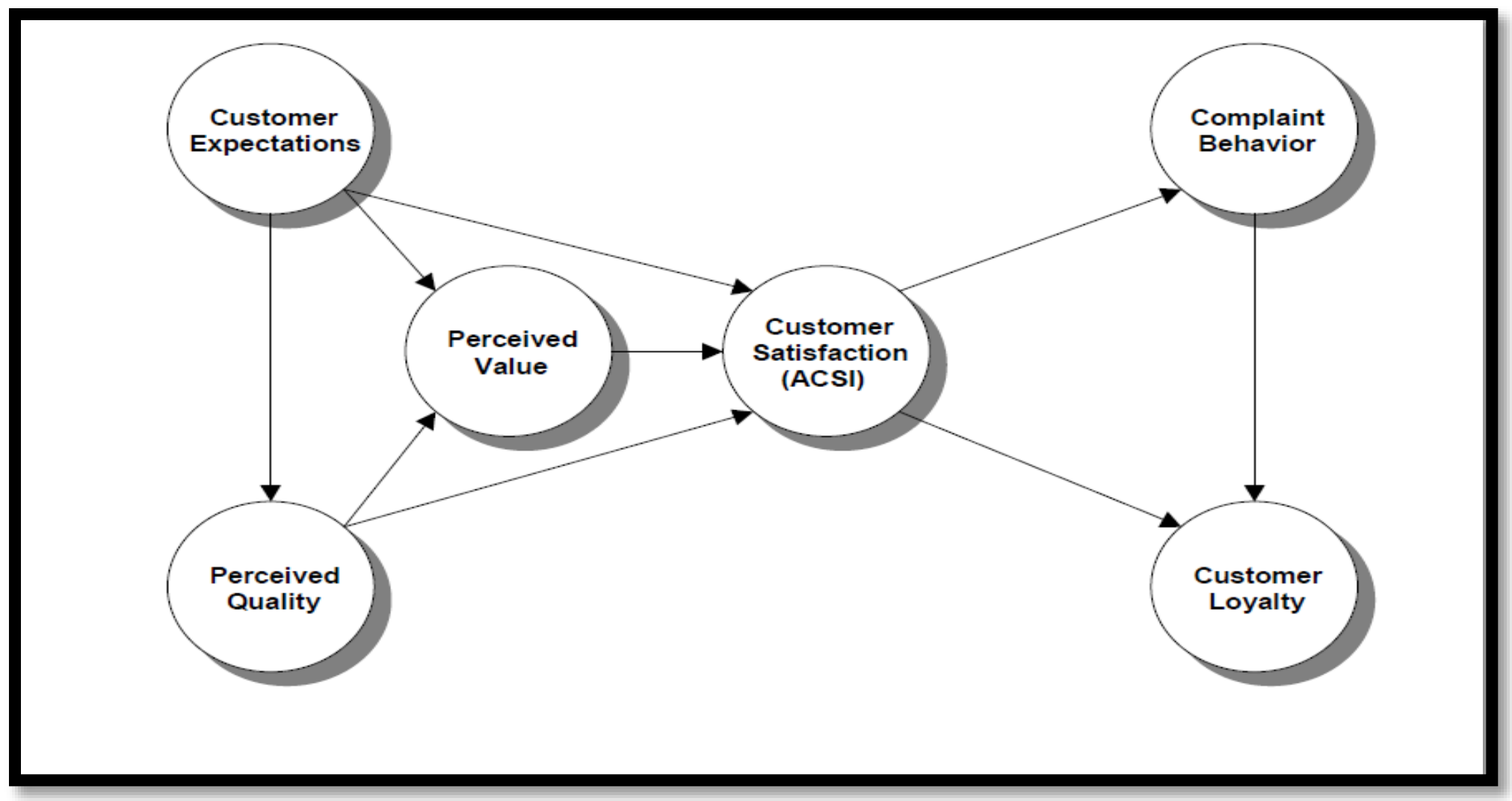

Figure 2.2 ACSI Model of Satisfaction

Furthermore, organizations should develop performance indicators to measure product or service success. Maidique and Zirger (1985) defined product or service success as, “the achievement of something desired, planned or attempted. While financial return is one of the easily quantifiable industrial performance yardsticks, it is far from the only important one. New product 'failure' can result in other important byproducts, organizational, technical and market developments." Cooper and Kleinschmidt categorized product performance measurements into financial performance, market impact and opportunity window. Furthermore, they implied the financial success of the product depends on those factors (Cooper and Kleinschmidt, 1987). Later Susan Hart suggested the measurement of product success can be divided into two groups, financial and non-financial. She then classified financial measurements into five groups, labeled as profit, assets, sale, capital and equity.

Many studies illustrate the impact of Lean Sigma in customer satisfaction. Based on the study conducted by Hekmatpanah, Shahin and Ravichandran, customer satisfaction has increased from 66.3% to 94% by deploying Six Sigma into the Sepahan Oil Company (Hekmatpanah, Shahin& Ravichandran, 2012). Lean Six Sigma provides the organization with methods and tools that guided them to identify and assess the firm competition ability to meet customer needs (Miskelly, 2012).

### 2.1.2 Impact of Customer satisfaction

As illustrated in previous sections, customer satisfaction has an enormous impact on the organization performances and is the key for company success. A considerable amount of literature has been published to describe the effect of customer satisfaction on organization performances. In general, the literature demonstrates its impact on three areas:

- Customer retention
- Customer loyalty
- Impact on shareholders

## 2.2. Quality concept:

The concept of quality has been changed dramatically in the last few decades. Based on the traditional concept of quality, it can be defined as the degree to which the product or service meet the standards. This concept ensures the product or service is free of negative values. In other words, the concepts ensure there is no unsatisfied customer. It doesn't add value to product or service to gain a complete advantage. Providing the customer with error-free service doesn't provide the company with loyal customers. The modern concept of quality is not only providing the customer with product or service free of negative value, but also exciting them and meeting their expectations. Many tools have been developed for that purpose such as QFD, Six Sigma, Voice of Customer, Kano model, etc. In the next part Six Sigma will be described in detail.

### 2.2.1. Six Sigma

Six Sigma was founded by the Motorola Corporation in the 1980s. Deploying Six Sigma in the Motorola Corporation resulted in a major improvement on its performance, saving about $16 billion. Later, many companies such as GE, Ford, General Motors and Honeywell adopted Six Sigma. Studies show Six Sigma has many positive effects on business performance. Those effects have a positive impact on both operational and financial performances. For example, GE saved $6.6 billion in the year 2000 by deploying Six Sigma. From a business perspective, Six Sigma can be defined as, a business improvement methodology which is used to improve profitability and effectiveness of the operations and to reduce quality costs and waste (Antony & Banuelas, 2001). It can also be defined as, "a business process that allows companies to drastically improve their bottom line by designing and monitoring everyday business activities in a way that minimizes waste and resources while increasing customer satisfaction" (Harry & Schroder, 2005). From the statistical viewpoint, it is considered a tool to reduce the variation in process whereas Sigma is a Greek word (σ) referring to standard deviation. Based on normal distribution, about 68.27% of the data falls in 1 standard deviation from the mean. Nearly 99.7% of the data falls 3 Sigma from the mean. That means if the company uses 1 Sigma, the number of defect would be 691462 per million. On the other hand, the goal of Six Sigma is to reduce the number of defects to 3.4 defect per million. Based on the above, we conclude that the concept of Six Sigma improves the business performance by reducing the number of defects and variation on the system (from the statistical viewpoint). Furthermore, Six Sigma provides companies with six major benefits: generates sustained success, sets performance goals for everyone, enhances values to customers, accelerates the rate of improvement, promotes learning and cross pollination, and executes strategic change (Pande, Neuman & Cavangh, 2000). In the following part the author will discuss Six Sigma methodologies.

#### 2.2.1.1 Six Sigma methodologies:

There are two types of six sigma methodologies: DMAIC and DMADV. DMAIC methodology is used to control and to improve the products or systems that already exist. DMADV is used to design and to develop the products or systems that either do not exist or are in the process of design.

##### 2.2.1.1.1 DMAIC:

DMAIC stands for define, measure, analyze, improve and control which is a five-phase process improvement methodology. The power of DMAIC comes from deploying many powerful quality tools such as cause and effect diagram, check sheet, histogram, control chart, Pareto chart, scatter diagram, Design of experiment, etc. Many authors and scientists argue that DMAIC has evolved from Deming's PDCA. DMAIC is considered an effective problem solving framework which is implemented in many organizations. DMAIC framework can deploy in almost all organizations, and it is not limited to Six Sigma. DMAIC's five phases are:

#### 2.2.1.1.1.1 Define

During this phase the team members are selected, and responsibilities and tasks for each team member are delegated. However, selecting the right person for the right position is not easy. At the same time, it is critical for project success. The Define phase has three steps. The first step is to initiate a project goal. The project charter is considered the most important tool in this step. The main role of this tool is to clarify project direction by identifying team roles and responsibility, setting up project's goals and identifying project objectives and constraints. Also, customers' critical to quality (CTQ) should be included in the project charter. SIPOC and value steam mapping are frequently used in defining process. SIPOC is high-level process map which stands for supply, input, process and customer, which are used to identify process elements. The fundamental difference between SIPOC and Value stream is that SIPOC is used to study a specific process while value stream is used to study the whole system (Chapman, 2012). Identifying customer requirements is a critical step to achieving customer satisfaction. Based on that, all customer requirements must be addressed carefully. Many tools have been developed to identify customer needs such as Kano Model, customer satisfaction survey and interviews.

#### 2.2.1.1.1.2 Measure:

The purpose of the measure phase is to identify critical values to the customer and measure performance of the system to customer requirements by going out and gathering data. Graphical analysis of variation tools such as Pareto chart and time-series are used in this phase. This phase consists of three steps. The first step is identifying the key process variable – input, output and process variables are carefully identified and measured at this step. Secondly, a proper data collection plan is established. Data collection method, data source, sample size and data type should be addressed at this step. The data collection plan must be reliable and reflect that the real system has been studied. Graphical tools give us good insight into how the process changes over time.

2.2.1.1.1.3 Analyze:

The main purpose of this phase is to pinpoint the root causes of process problem and inefficiency. During this phase, theory is developed to test the data that has been collected in previous phases where in-depth study should be conducted. Many statistical and quality tools such as hypothesis testing, brainstorming, anova and fishbone diagram are deployed in this phase. At the end of this phase, we will get a list of verified root causes to our issue.

2.2.1.1.1.4 Improve:

The general gained knowledge from measure and analyze phases is used in this phase to generate optimal solutions which lead to improved system performances in return. During this phase, barnstorming, benchmarking, TRIZ and Design of Experiment (DOE) is implemented to find the best solutions and prioritize root causes as related to customer requirements, identified in the define phase. Proposed solutions should be tested in order to acquire validated solutions.

2.2.1.1.1.5 Control:

Since the control phase is the final phase, the team should ensure all solutions are implemented in full scale and all incomplete work must be finished properly. The main purpose of this step is to ensure the system stays stable over time. This phase helps create a sustainable system. SPC is the most commonly used tool in this phase, employed to monitor the change in system performance over time. However, there are many tools and techniques that may be used such as mistake proofing, FMEA, TPM and control plan.

### 2.2.2 Lean Six Sigma

Nowadays, most organizations use the Lean Six Sigma approach because it provides the organization with powerful methodologies to reduce both waste and variation. Lean Six Sigma was first used in 1997 by BAE system. Lean Six Sigma can be viewed as a creative combination between Lean Thinking and Six Sigma. Lean Thinking is derived from Toyota Production System (TPS) which simply means eliminating waste from the process or system. According to Womack and Jones, the Lean concept has five phases. The first phase is represented by identifying the customer and specifying the value. At this step the team should identify two types of values. The first type is called value added activities which represent all activity customers are willing to pay for. The second type is the non-added value activity which can be divided into two types. The first type is non-added value necessary wastes which represent all activities that the customers are not willing to pay for, but those activities are important and cannot be eliminated. The other type is non-added value pure waste which represents all activities that don't add values to product and can be eliminated such as rework, inventory. The second phase of the five, is identifying value stream, at which all the steps that are required to create the value are highlighted. Based on Womack and Jones, value stream can be defined as "set of all specific actions required to bring a specific product through three management tasks: problem solving task, information management task and physical transformation task." Identifying value is an important step that helps us to manage scope, prevent confusion and eliminate waste. The third phase is flow, with the main purpose being to ensure the process is smooth and to avoid the peak, gap and disconnecting in the process. Pull is the fourth phase, which simply means to produce only if there is demand. In return, the excess inventory and overproduction costs are reduced and the process speed is increased at the same time. The last phase is perfection through elimination of waste.

## 2.3 Service quality

Today, service quality becomes the central focus of many organizations worldwide. Due to increasing the competition, organizations become more concerned about measuring their service quality and more motivated to improving them. Unfortunately, measuring service quality is not as simple as measuring product quality because most service characteristics are intangible. Before going further, it is important to define the service. Gronroos defined the service as:

> "Activity or series of activities of more or less intangible nature that normally, but not necessarily, take place in interaction between the customer and the service employees and or physical resources or goods and/or systems of service provider, which are provided as solutions to customer problems" (Gronroos, 1990).

Service is also defined as, "any intangible benefit, which is paid for directly or indirectly, and which often includes a larger or smaller physical or technical component" (Andresen, 1983). Both definitions illustrate direct interactions between the customers and the firm which reveal the fundamental difference between services and goods. In 1985, Zeithmal, Parasuraman, and Berry stated that the service is different from goods in three aspects:

- Service is intangible
- Service is heterogeneous which reflects the service variety from producer to other.

- Producing and consuming the service are correlated.

For decades, an enormous number of studies have been conducted to study the quality concept in service industry. Service quality is defined as a disciplinary level between clients' expectations or needs and their perceptions (Zeithmal, Parasuraman, and Berry, 1990). In other words, it represents the difference between customer perception and expectation which could be either positive or negative. Gronroos claimed that service quality depends on two dimensions – functional and technical (Gronroos, 1990). Furthermore, Gronroos described the functional dimension as what a customer gets from the service. The technical dimension represents how the service is delivered. In 1985, Zeithmal, Parasuraman, and Berry suggested the service quality is based on ten dimensions. Later, they narrowed it to five dimensions. In the next section, those dimensions will be described in detail.

## 2.4. Customer satisfaction gaps

In 1985, Zeithamal, Parasuraman, and Berry developed a customer gap analysis model. Five gaps have been identified as the main reasons for service quality problems.

First is the communication gap, which represents the difference between promises made through advertisement or other types of communication and the actual service or product. This gap leads to vast customer dissatisfactions because expectations will not be fulfilled (Hill & Alexander, 2006). The major cause for this gap is the company's marketing communications. By making promises and creating unrealistic values in the customers' mind – which are hard to fulfill, leading to disappointment – they lose trust and respect for the company.

Second is the knowledge gap, which models the gap between managers' perceptions and customer expectations. This gap can be traced back to inaccuracy in identifying customers' needs and requirements. To reduce this gap, service features and level of performance for each feature should be identified completely, prior to releasing the service.

Third is the policy gap, which represents the difference between management perception and service quality specification. Incorrect translation of customer's expectation into appropriate operating procedures is the main reason for this gap. This, in return, will result in poor satisfaction with product and services even though management has a full understanding of customer requirements.

The fourth kind of gap is delivery, exhibiting the gap between service quality specification and service delivery. This gap can be traced back to weakness in employee performance. Even though the organization has clear procedures to match customer needs, it will still fail to deliver customer satisfaction if it doesn't have well-trained employees to deliver its values to customers. To solve this problem, the company must develop a training program to ensure that employees follow procedures and deliver what the customer expects.

The final gap is the customer gap. This gap occurs when customer perception is different from delivered product or services. The SERVQUAL instrument has been constructed to measure gap five. SERVQUAL represents the service quality and is influenced by the previous four gaps. Customer satisfaction with the service delivered, depends on this gap.

## 2.5. Models and instruments to measure and analyze Customer satisfaction

### 2.5.1. SERVQUAL

In 1988, Zeithamal, Parasuraman, and Berry developed 22 items to measure the quality of provided services, which is called SERVQUAL. After studying the customer behavior in different service sectors, they uncovered 10 dimensions of quality: responsiveness, reliability, tangible, competence, courtesy, credibility, security access, communication and understanding (Oliver, 1997). Afterwards, those dimensions were primarily used to evaluate the service in many organizations.

SERVQUAL was constructed based on two parts, expectation and perception. The first part, expectation, is specified to capture the desired and needed level in the service from the customer viewpoint. Perception reflects the customer judgment regarding delivered services. A 22 item survey was developed to evaluate each area by asking the customer to rate each item on a seven-point Likert scale. Then, the gap between the perceived and expected level is achieved by subtracting the expected level from perceived level. A positive gap means the provided service is good and the customer is satisfied, and vice-versa. Many statistical tools are used to analyze these surveys.

Afterward, the same authors indicated that the SERVQUAL five dimensions can be used effectively to capture the original 10 dimensions of service quality (Zeithamal, Parasuraman & Berry, 1990). The five dimensions are: tangibles, reliability, responsiveness, assurance and empathy. In other words, the customer satisfaction in service industry depends on those five dimensions (see Figure 2.3). The importance of each dimension is not equal and varies from service-to-service. However, many studies show that reliability has the highest weight score among the five dimensions in many service industries. In other words, the reliability dimension has the highest impact on customer satisfaction. The definitions stated by Zeithamal, Parasuraman, & Berry for each of those dimensions are listed below:

- Tangibles represent the physical appearance of the facility, communication material, and employees appearances.
- Reliability reflects the capability of the service provider to carry out the promise service dependably and accurately.
- Responsiveness represents the organization willingness to serve their customer and provide them with promoted services.
- Assurance encompasses employees' knowledge, courtesy and their ability to win customer trust and confidence.
- Empathy exhibits both the individual attention and caring that the organizations provide their customers.

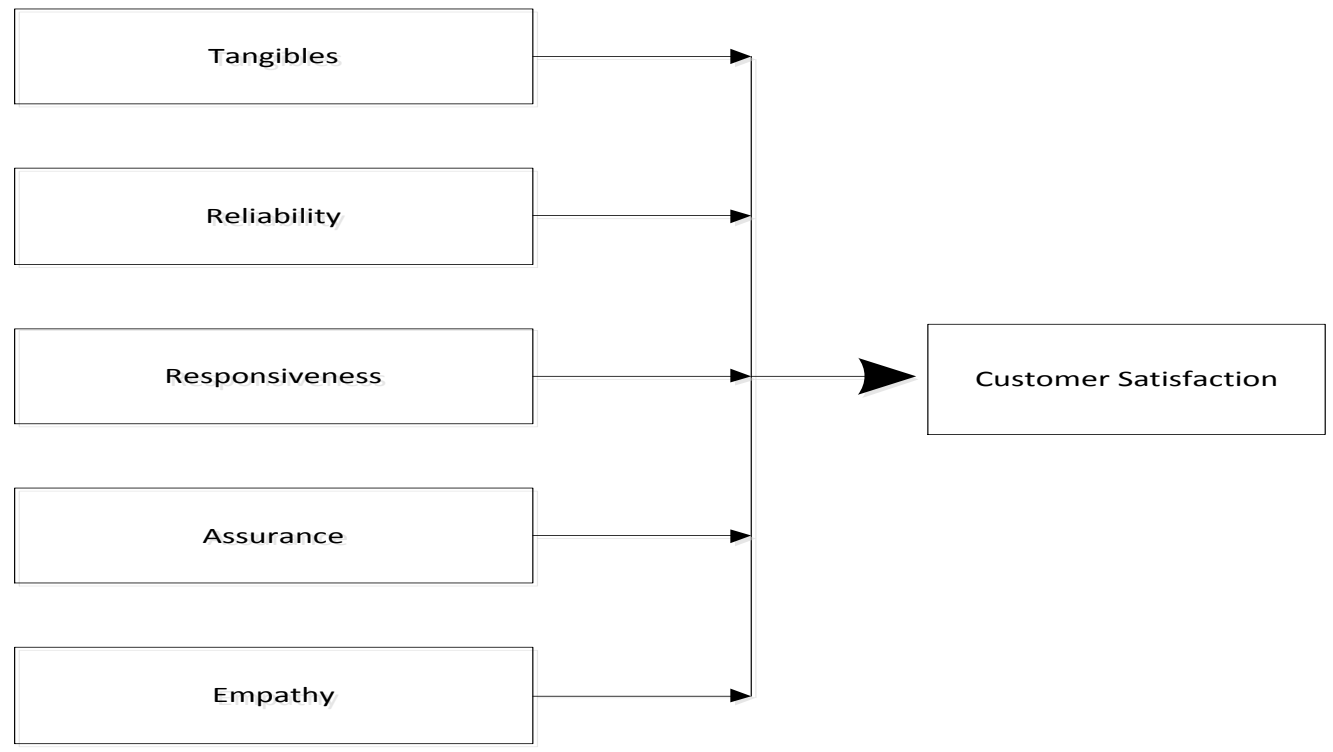

Figure 2.3 Impact Dimension of Service Quality on Customer Satisfaction

In spite of the critiques for SERVQUAL by many literature reviews, it is still the most widely used instrument to measure and analyze customer satisfaction. Ladhari, stated that SERVQUAL is an effective tool for service companies to distinguish themselves over their competitors (Ladhari, 2008). Furthermore, Ghylin mentioned that SERVQUAL aids the firms to improve their performances and deliver high-quality services (Ghylin, 2008). In conclusion, many publications demonstrate the importance and effectiveness of SERVQUAL to measure the customer satisfaction which is the main reason for choosing it in this study.

### 2.5.2. Kano Model:

In 1984, Noriaki Kano developed a model to identify and clarify the customer requirements and needs which is known as the Kano model. He assumed the customer thinking in two dimensions, so he graphed these key quality characteristics with a horizontal axis whereas the higher level of quality will be close to the right and the lower level to the left (Chen, Chang & Huang, 2009). Kano categorized the product attributes into four types (Kano, 1984):

1. The Must-be attributes (also known as Basic) refers to the product or service characteristics that are expected by the customer, which leads to a massive dissatisfaction if they are absent or poorly satisfied (Xu, Jiao,Yang, Khalid, Opperud, 2008). Fulfilling those quality attributes will not excite the customers. For example, having a good brake system in the car will not increase customer satisfaction. On the other hand, an inefficient brake system will make the customer extremely dissatisfied.

2. The one-dimension attributes represent those attributes that will increase customer satisfaction if they are satisfied. Otherwise, these attributes will lead to dissatisfaction. A higher level of fulfillment leads to higher customer satisfaction and vice-versa. For example, in automobiles the gas mileage is one-dimension. If the car has better gas mileage, customer satisfaction is increased and vice-versa.

3. Excitement attributes (also called wow!) are those needs which the customer does not know. Satisfying those needs will have a major impact on customer satisfaction. Since the customer is not aware of those needs, the failure to meet them will not have any influence on customer satisfaction.

4. Indifferent needs: those needs that the customer has that do not affect the level of satisfaction, and they do not result in customer dissatisfaction if they are not fulfilled. Sauerwein, Bailom, Matzler, and Hinterhuber summarized the major advantages of the Kano model: (1) Assist the designer to priority for product development. (2) Help the company to identify product requirement. (3) Assist the organization to adapt process oriented for product development. (4) Provide valuable assistance in product or service development.

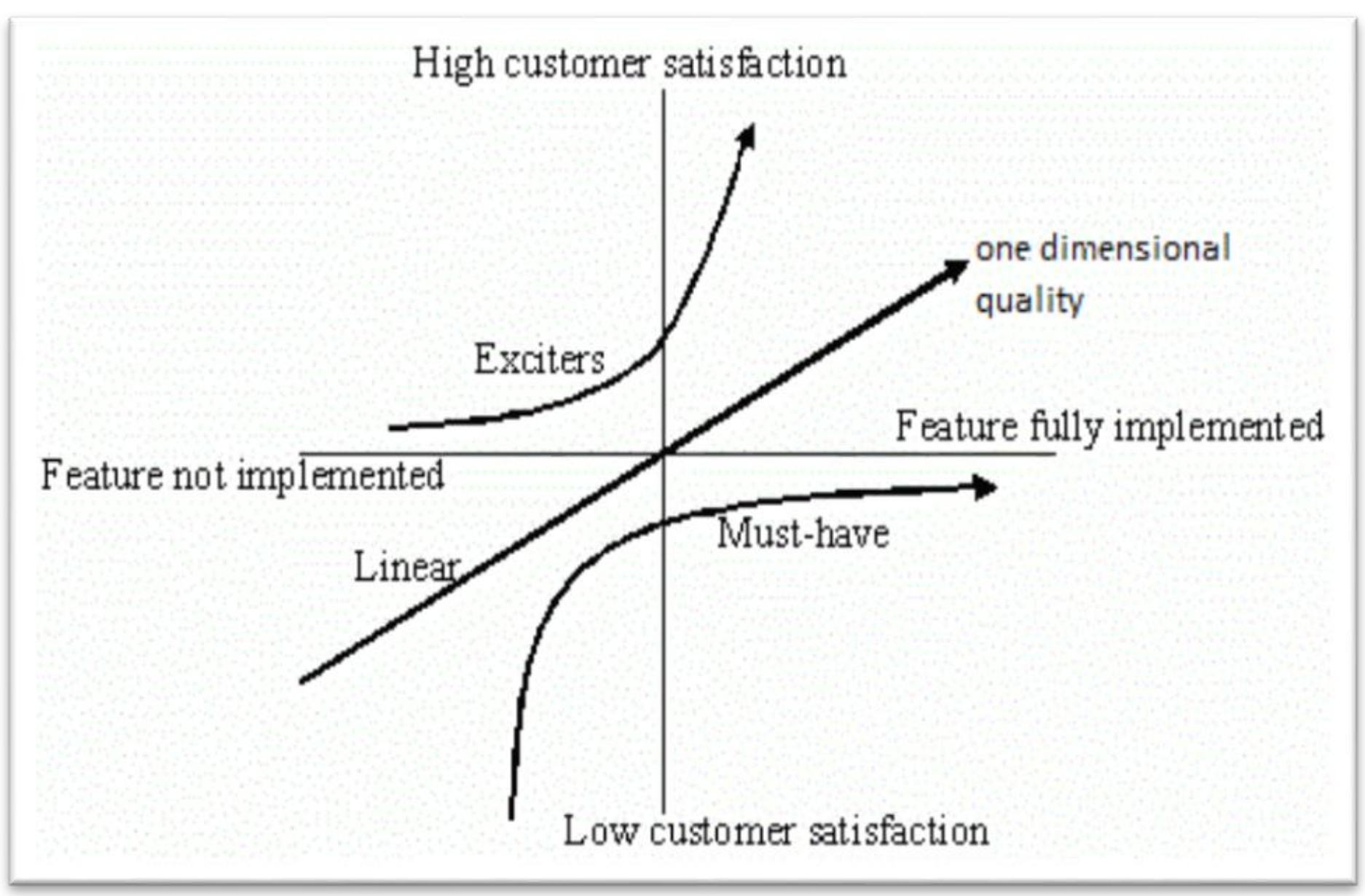

Figure 2.4 Kano Model (Kano, 1984)

<u>2.5.2.1 Integrating Six Sigma with Kano model:</u>

A model to integrate Six Sigma with Kano model has been developed by Chen, Chang & Huang. Their framework is specified to identify the key quality characteristics from the customer perspective by using both Kano model Six Sigma (DMAIC) as an optimization tool. This framework can be summarized in the following steps: Defining the level of expectation before using and the level of satisfaction after using is the first step. Three index values for evaluation should be created to determine the level of satisfaction and expectation. After that, the achieved customer satisfaction with the product or service must be measured by using the questionnaires method. Later, each index will be analyzed separately to develop the measurement matrix for both the arrived and the satisfaction level. Kano would be constructed by integrating both the vertical distance between the coordinate and diagonal points. The last step is Improve and Control. Exceeding customer expectation above exceptional level should be considered to improve satisfaction or reduce dissatisfaction. Then, improvement for above the exceptional level must be made. Lastly, the Kano two-dimensional model should be reconstructed by the questionnaires method to confirm the results of improvement. If we get poor results from a previous step, a corrective action must be considered to develop a new solution (Chen, Chang & Huang).

# CHAPTER 3
# RESEARCH METHODOLOGY

## 3.1. Introduction

Research methodology can be defined as a scientific methodology for solving the problem and generating a new knowledge for specific puzzles in general. The author has composed this chapter to explore the research strategy that has been used in this study. In the first part, the research design, methodology and questions have been described along with the data collection method. The main goals and the choice of study have been investigated and pinpointed in the next part. In the final part, the main reasons for conducting this study have been explained.

In general, this study has been conducted to develop a comprehensive framework to measure and analyze customer satisfaction. The framework power comes from using effective tools and methodologies which are Six Sigma, SERVQUAL and QFD. SERVQUAL instrument has been used to identify the customer requirements and highlight the gap between perceived and expected levels. Six Sigma contributes effective tools to analyze the customer needs and requirements. By analyzing and eliminating the root causes of customer dissatisfactions, the companies will boost their profits. Later, this framework will be deployed to measure and analyze customer satisfaction in XYZ Company in Iraq.

## 3.2. Research design:

This study has been constructed based on two streams. The first part is designed to develop framework to measure and analyze customer satisfaction – the satisfaction key drivers will be identified and explained. The second part is structured to study and analyze customer satisfaction in computer maintenance companies. Based on Bryman and Bell, the research design can be divided into five categories: comparative, cross sectional, case study, longitudinal and experimental. Since case study design provides the researcher with in-depth and close examination of the problem, it has been chosen to answer the problem formulation in part two. The case study is designed to study the customer's attitude regarding computer maintenance service. Two types of surveys have been developed for this purpose. Moreover, these surveys are intended to capture customer expected and perceived level of satisfaction. Later, those surveys (the collected data) are analyzed by the framework that has been developed in section one. This study has been completed in eight steps which are:

1. Select the area of study: The author of this study has both the experience and a great interest in Lean Six Sigma. For this reason this area has been selected.
2. Review the literature: Enormous numbers of publications have been reviewed by the author to discover the gap in the knowledge.
3. Select the subject: Based on reviewing the literature, the author discovered no study has been made to measure the customer satisfaction by Lean Six Sigma. Moreover, the customer satisfaction has become essential in global marketing. Based on the previous two reasons the subject of study has been selected.

4. Identify research questions: This study uncovered five fundamental questions which must be answered. Section 3.2.2 has been developed to explain those questions.
5. Define research objectives: In this step, two main goals have been stated: The first objective constructs a framework to measure and analyze customer satisfaction. The second defines the dimensions of service quality in computer service industry.
6. Developing Framework: An extensive study has been made by the author to develop a robust framework to measure the customer satisfaction in the computer service industry.
7. Case Study: To answer questions three, four and five and to get an in-depth understanding of the problem formulation, this case study has been selected. This step is subdivided into three steps: questionnaire design, data collection and data analysis.
8. Draw a conclusion: Based on the previous steps, the root causes of the problems have been identified and the conclusion has been made.

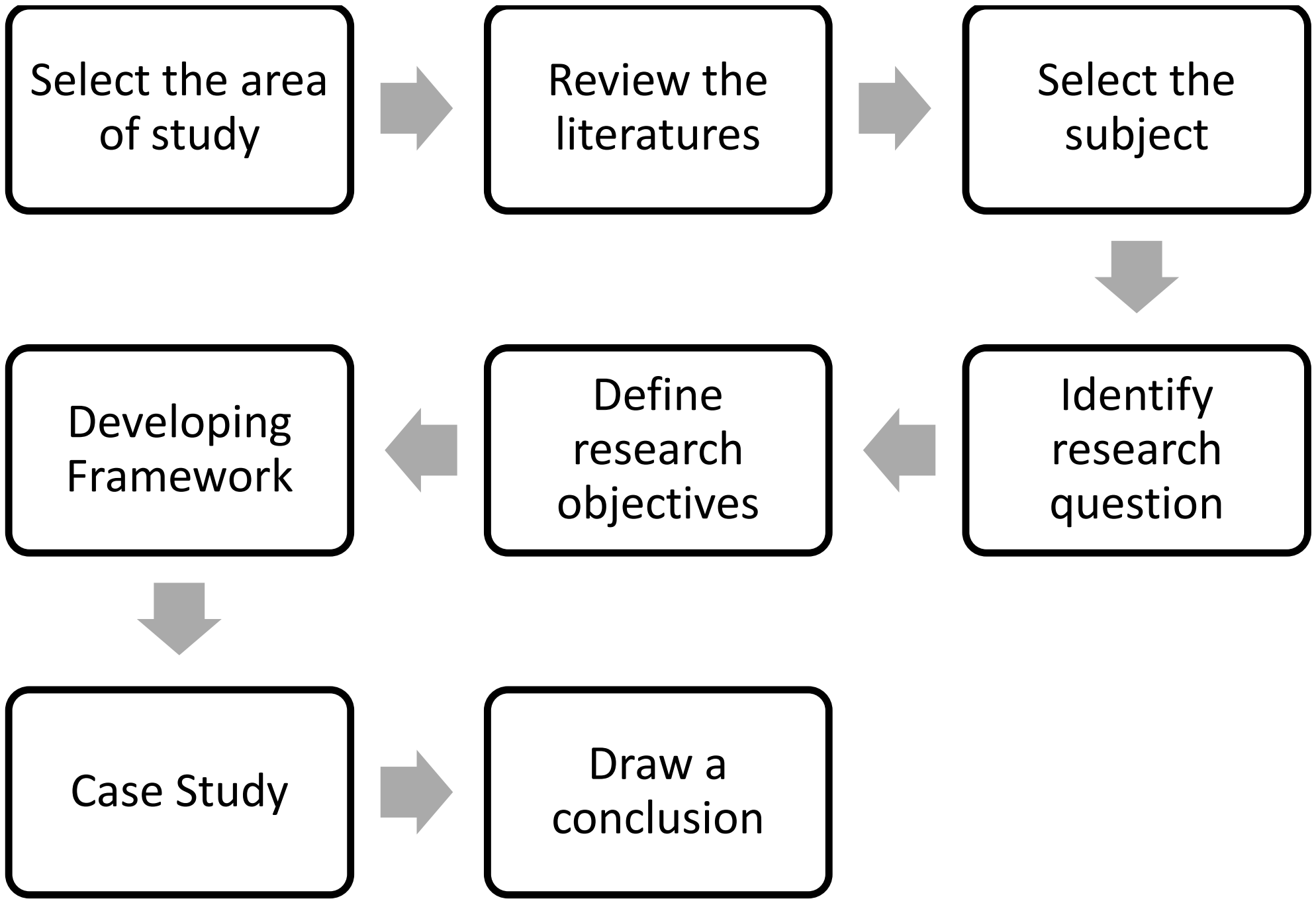

Figure 3.1 Research Framework

### 3.2.1 Research method:

Quantitative and qualitative are the most widely used strategies in research. Each method has advantages and disadvantages and no method is better than the other; however, the selection of the method depends on the nature of the study. Based on Creswell, Qualitative research can be defined as "an inquiry process of understanding based on distinct methodological traditions of inquiry that explore a social or human problem. The researcher builds a complex, holistic picture, analyzes words, reports detailed views of informants and conducts the study in a natural setting" (Creswell, 1994). On the other hand, quantitative strategy is aimed to test the theory and to look at the cause and effect. It can be defined as, "a means for testing objective theories by examining the relationship among variables. These variables, in turn, can be measured, typically on instruments, so that numbered data can be analyzed using statistical procedures" (Creswell, 2008). Quantitative method has been adapted in this study. The fact that the quantitative method is preferred over the qualitative method is because of compatibility to the research questions. Furthermore, this method is adequate for measuring the customer satisfaction key drivers and assessing the differences between the customers' perceptions.

### 3.2.2 Research questions:

The main purpose of this thesis emerged from the following questions:

1. How do computer service companies deploy Lean Six Sigma to measure and analyze customer satisfaction effectively?
2. What are the key quality characteristics or dimensions of service qualities of computer service?
3. How does XYZ Company utilize and improve customer satisfaction by using the framework that has been developed in this study?
4. What is the impact of each dimension in the overall customer satisfaction at XYZ Company?
5. What are the root causes of customer dissatisfaction at XYZ Company?

### 3.2.3 Questionnaire Design:

Since the survey has been conducted in Iraq the questionnaire is written in English first and translated to Arabic later. Based on the customer satisfaction's key drivers and SERVQUAL, the questionnaires have been constructed. In general, the formulated questionnaire can be spilt up into two distinct parts. The first part is designed to capture the voice of the customer and to measure the customer expected level. In this part, the author seeks answers about how the state of Computer Maintenance services should look. The second part, is aimed to measure customer perceptions. In return, it assists us to rate the organization's performances in regard to customer expectation. Appendix B and C illustrates those questionnaires.

#### 3.2.4 Data Collection

Since this study has been designed to capture both customer expectations and customer perceptions, two types of surveys were developed for that purpose. A 42 item questionnaire has been developed to capture both customer expectations and customer perceptions. Two types of data are collected; one type reflects customer expectations and the other type represents customer satisfaction. The surveys are designed to measure service quality at computer maintenance. The customers have been provided with a survey while waiting to receive their services. This research is limited to the current customers. To ensure that all participants are XYZ Customers, the survey papers are handed over to the customers after they have turned over their computer for repair. After completing those surveys, they are returned to the front-desk employees. The total observed customers is 164 (82 for each survey). The duration of an individual participant to complete one survey is 6 minutes. Afterward, the collected data is saved on an Excel spreadsheet. Later, this data is analyzed by SERVQUAL and different statistical tools such as Anova and descriptive statistics.

## 3.3 Study objectives

Building the right product and delivering the right service are fundamental business drivers. Many publications show that providing customers with high quality products is not enough if they are not accompanied with good customer care. By entering large computer technology organizations such as Dell and HP in the service market and developing the new online-based tools – used for problem diagnosing and repairing – the competitive pressure has been increased on small and medium-sized businesses. Based on that, this study has been conducted to provide the small and medium-size firms with guidelines to measure and analyze customer satisfaction of their delivered services. Investigating the root causes of customer dissatisfaction and identifying their impact on the overall customer satisfaction of those attributes will assist the organization to prioritize its efforts and improve the service weaknesses. This in return, increases customer satisfaction and promotes customer loyalty. Figure 3.2 illustrates the main objectives for this study.

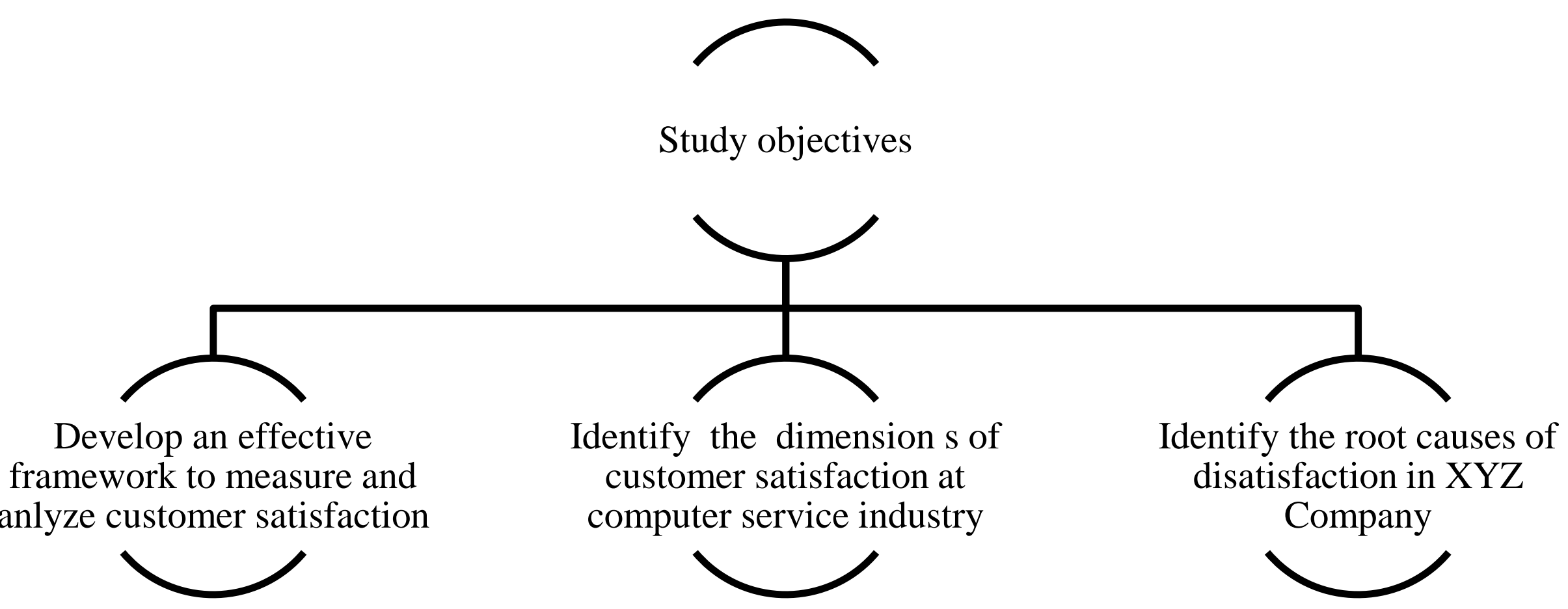

Figure 3.2 Study Objective

### 3.4 Need of the study:

As previously mentioned, increasing the competition makes many computer service companies powerless in facing large organizations. A number of studies show that computer service companies lose about 25 to 40 percent of their customers each year, and many of these companies do not know why, how or where they lose their customers. They do not have a well-designed system or strategy to measure and analyze customer satisfaction, which is considered the main reason for customer decline. Furthermore, there are some companies that achieve high customer satisfaction while other companies still struggle in that area.

XYZ Company is one of those companies who is suffering from losing customers continuously without knowing what makes their customers dissatisfied. XYZ Company has received numerous complaints, but they don't reflect the overall causes of customer dissatisfaction. Many studies show that not all dissatisfied customers express their feelings toward corporate managements, while other studies show they are most likely telling other customers which means the company not only loses the current, dissatisfied customers but future customers as well.

Many of XYZ Company administrators agree there are a number of unsatisfied customers, but they don't know what makes those customers unsatisfied, or what the level of satisfaction with the service is. The root causes with the service haven't been identified, which causes the mangers to struggle in developing an effective improvement plan. This study will provide XYZ Company – and computer service companies in general – with effective tools to improve customer satisfaction and increase their profits.

## 3.5. Study contributions

As stated in Chapter 2, Lean Six Sigma is the most powerful improvement methodology, deployed in almost all industrial sectors. However, no study has been conducted to explain the deployment of this tool to measure or analyze customer satisfaction. Considering customer satisfaction has become the key driver for business success. Based on that, this study has been developed to provide the companies with a structured approach to measure customer satisfaction by using both Lean Six Sigma and SERVQUAL instrument.

## 3.6. Limitation and scope of the study

The scope of this study is limited to study and analysis of customer behavior in Computer Maintenance Companies. In other words, this study is restricted to report the satisfaction level regarding the provided services. There are two major components, customer expectations and customer perceptions. These components contribute four major limitations, listed as follows:

- The data has been collected from one single computer service company, so the result may not represent the whole population.
- By nature, customer satisfaction surveys have a sort of variability since not all the customers will participate in this study.
- Due to time and available resources, a limited number of data has been collected.
- To increase the number of participants, the customers have asked to evaluate only the important key characteristics (questionnaires). For instance, 17 key drivers have been selected out of 28 key drivers, which represent the overall dimensions of customer satisfaction and computer service.

# CHAPTER 4
# FRAMEWORK TO MEASURE AND ANALYZE CUSTOMER SATISFACTION

## 4.1 Overview

Due to global competition, satisfying the customer has become the main goal for many organizations. To survive in a vital global market, the organization has to provide their customer not only with a good product or service, but also with good customer care. "The customer is always right" or "the customer first" become the most used slogans by many organizations.

Considering the massive growth in computing units, the necessity for computer services has increased. Based on the study conducted in 2008, the computer service companies reap 4.1 billion dollars. However, another study shows that many of those companies lose 30 to 40 percent of their customers each year; that can be traced back to the lack of a solid customer measurement system. This present study has been developed to integrate Lean Six Sigma with SERVQUAL to measure and analyze the customer satisfaction in computer service companies.

This chapter is specified to explore and explain the framework to measure the customer satisfaction and its tool. Furthermore, 28 questionnaire items have been developed to measure the customer satisfaction in different types of computer service companies. Those items are derived from the five dimensions of service quality. In summary, this chapter aims to answer questions one and two by providing the reader with a comprehensive clarification for deploying Lean Six Sigma, SERVQUAL and Kano in computer service companies to maximize customer satisfaction and increase customer loyalty.

## 4.2. Drivers of customer satisfaction:

The drivers of customer satisfaction represent all the elements that affect customer judgment for particular service or product as well as understanding those keys will help the firm to be conscious of what components are most important for the customers. Since they have an enormous impact on service performance, those drivers must be precisely identified prior to conducting the survey to evaluate their impact. Two kinds of questions must be asked to assess the influence of the key drivers about the satisfaction level. First the customer should be asked to rate the importance of those key characteristics and their impact on overall satisfaction. The second fold is specified to identify the future key drivers by asking the customers about what features are missing in the service. Many researchers argue there is a strong correlation between organization performance and those key drivers. However, many executives and corporate management do not have accurate understanding about what the key drivers of customer satisfaction are, that explains why some organizations do better than others. Those drivers vary from product to product and from service to service. QFD, Kano model, and many quality tools are used to pinpoint them. By studying many companies' models and reviewing many literatures, the author proposes five groups of key drivers that drive customer satisfaction across many computer service organizations. However, most of these key drivers are broken down into subcategory drivers.

### 4.2.1. Reliability

Reliability is one of the most critical factors that help the firm to differentiate itself from the others and increase the satisfaction. It represents the ability of the service provider to deliver the promised service right the first time. To provide the customer with a reliable service, three conditions must be accomplished. First, the expected level of service performance must be fulfilled. In other words, the service quality should meet or exceed the customer expected level. Next, the service should be available to use whenever the customer wants. The last condition is the continuity. The previous condition revealed three key drivers which must be optimized in computer services. Firstly, the firm must ensure that its customers receive their service (computer repair, data backup & recovery, spyware/virus removal, security services, server installation and support) at the promised time. Subsequently, those computers have personal or private company information, therefore the firm must ensure the employees respect customer privacy, and understand the importance of safekeeping the information. Ensuring the customer receives decent care and the charges are comparable with the provided service is considered a critical key characteristic since they impact the company credibility and lead to enormous customer dissatisfaction. To sum up, the reliability dimension has three key drivers which are labeled commitment, security, and performance.

Table 4.1 Reliability Key Drivers

| Key driver | The questionnaire |
|---|---|
| **Service performance** | The customers should be asked if they received a complete care from the organization. |
| | The customers should be asked if the company insisted on error free service. |
| | The customers should be asked if they received the service right at the first time. |
| | The customers should be asked if the technicians showed sincere interest while solving their problem. |
| **Time line of service** | The customers should be asked if they received service at the promised time. |
| **Service security** | The customers should be asked if the technicians respect their privacy. |

### 4.2.2. Tangibles

Based on Zeithamal, Parasuraman, and Berry, tangibles encompass all physical aspects of service like physical facility, equipment and personal appearance. This reveals three key drivers in computer service. The convenience of waiting place is one key. Many studies show that the customers spend on average 20 to 40 minutes before they get the service, so the firm must ensure that these areas are clean and comfortable and have enough entertainment. However, many organizations ignore the importance of this key whereas many studies estimate the impact of this factor between 5-8% overall satisfaction. The second factor is the appearance of employees and equipment. This factor has a moderate impact on customer satisfaction. Since many customers are unwilling to pay for this factor, many researchers consider it as non-added value. The convenience of the procedure (paperwork) is the other key which must be simple and it takes a short time. A considerable number of publications show the complicated procedure (paperwork) inconveniences the customer and in return increases the dissatisfaction.

Table 4.2 Tangibility Key Drivers

| Key driver | The questionnaire |
|---|---|
| **Appearance of employees and equipment** | The customers should be asked if the employees dressed professionally. |
| | The customers should be asked if the equipment is modern looking |
| **Waiting area conformity** | The customers should be asked if the waiting area is clean. |
| | The customers should be asked if the waiting area is comfortable. |
| | The customers should be asked if the waiting area had enough entertainment. |
| **convenience of the procedure(paper work)** | The customers should be asked if the repair order is easy to fill. |
| | The customers should be asked if it takes a short time to fill the repair order out. |

### 4.2.3. Empathy

Empathy represents the company's intention to value and take care of the individuals. Zeithamal, Parasuraman, and Berry argue that empathy can be divided into three factors. First factor is accessibility which represents the customer ability to contact the organization whenever s/he uses or encounters a problem. Communication is the second factor which demonstrates the firm ability to keep their customers informed in the language that they can understand. Last, understanding the customer factor exhibits the employees' effort to understand customer needs and to provide them with special attention and recognition.

Table 4.3 Empathy Key Drivers

| Key driver | The questionnaire |
|---|---|
| **Accessibility** | The customers should be asked if the company operation hour is convenient to the customer. |
| | The customers should be asked if the parking lot is near to the service location. |
| **Communication** | The customers should be asked if the employees give the customer individual attention. |
| | The customers should be asked if the employees use a simple language while they communicate with you. |
| **Understanding** | The customers should be asked if the employees understand the customer specific needs. |

#### 4.2.4. Responsiveness

Responsiveness represents the readiness and willingness of the employees and the firm to help and provide the customer with service whenever needed, for instance resolving the customer problem instantly. There are four key drivers associated with responsiveness uncovered in this study. The first key represents the employees' speed of the response to the customer. Clark, Kaminski, and Rink point out that the quick response will increase the satisfaction with the service. However, a quick response is not enough if it is not accompanied with a professional handling of problems and complaints. Employees' attitude toward the customer is another key. Harter, Schmidt, and Hayes (2002) studied 7939 business units in 36 different industries, and they found a strong correlation between employees' attitudes and customers' satisfaction. The last key is the availability of the employees. They should never be too busy to respond to service requests. The following table illustrates the key drivers and the questions that define them:

Table 4.4 Responsiveness Key Drivers

| Key driver | Questionnaire |
|---|---|
| **speed** | The customers should be asked if the employees responded to them quickly. |
| **Accuracy** | The customers should be asked if the employees told them exactly when they would receive their service. |
| **Availability** | The customers should be asked if the employees were too busy to help them. |
| **Attitude** | The customers should be asked if employees showed a positive attitude when they apologized. |

### 4.2.5 Assurance

Zeithamal, Parasuraman, and Berry define assurance as the employees' competence, knowledge and courtesy, and their ability to win customer trust and convey confidence. In many cases the firm provides the customers with excellent service, yet the customers are still unsatisfied because they feel that the employees do not provide the desired attention. Therefore, the customer's judgment about the service will be impacted negatively. For example, the employees may install the server and do their job completely, but they are not cordial enough nor do they ask the customer if there is anything else they want. This affects the customer's judgment in return. There are four key drivers associated with assurance. Primarily, the employee's courtesy reflects politeness, respect and friendliness towards the customer. Many studies show that even though the customers have a negative attitude regarding the service, their level of loyalty may increase if the employees on the front line are polite, friendly and handle their problem professionally with courtesy. Credibility is another driver which represents the company's honesty in dealing with their customer. Many companies realize the importance of this factor by not breaking their promises and keeping customer loyalty. The competence of employees is considered a critical key for service quality. It represents the knowledge and skill of employees, which is required for delivering the service. Lastly, the security driver is represented by delivering the customer free risk service, and keeping their data and financial information secure.

Table 4.5 Assurance Key Drivers

| Key driver | The questionnaire |
|---|---|
| **Courtesy** | The customers should be asked if the employees were courteous |
| | The customers should be asked if the technician were courteous |
| **Credibility** | The customers should be asked if the cost of repair is reasonable |
| **Competences** | The customers should be asked if the staff are knowledgeable |
| **Security** | The customers should be asked if they feel their personal/organizational information are secure |
| | The customers should be asked if they feel their payment information is secure. |

4.3 Framework to measure and analyze customer satisfaction:

Due to the global competition, many organizations realize the importance of customer satisfaction for their long term survival. Without question, satisfying the customer becomes the main operational goal for almost all organizations all over the world; however, many organizations have failed to sustain their customer's loyalty even though they comprehend its importance for overall success. This uncovers a fundamental question as to why a number of organizations still aggressively struggle to satisfy and keep their customers while others do not. A considerable amount of research has been undertaken to answer this question. Numerous researchers justified that lacking a proper satisfaction measurement system is the main reason for customer decay. Based on previous fact, the author decided to develop a framework to help the organization measure and analyze customer satisfaction.

This framework brings together the key aspects of Six Sigma and SERVQUAL instrument into a structured approach to measure and analyze customer satisfaction in computer services. It deploys the DMAIC problem solving methodology to provide the organization with a good insight about what the customers' needs and requirements are, along with SERVQUAL, which contributes service dimensions and the Gaps Analyze technique. It can be summarized in nine steps which will be discussed in the following part. Figure 4.1 illustrates those steps and Figure 4.2 illustrates the tools that have been used in the framework. The major framework steps are as follows:

- Define the process variables
- Identify the voice of customer
- Categorize customer requirements by Kano model
- Translate voice of customer into critical to quality
- Measure satisfaction level:
- Define data collection system
- Validate the collected data
- Analyze the Gap by SERVQUAL
- Analyze the root cause of customer dissatisfaction

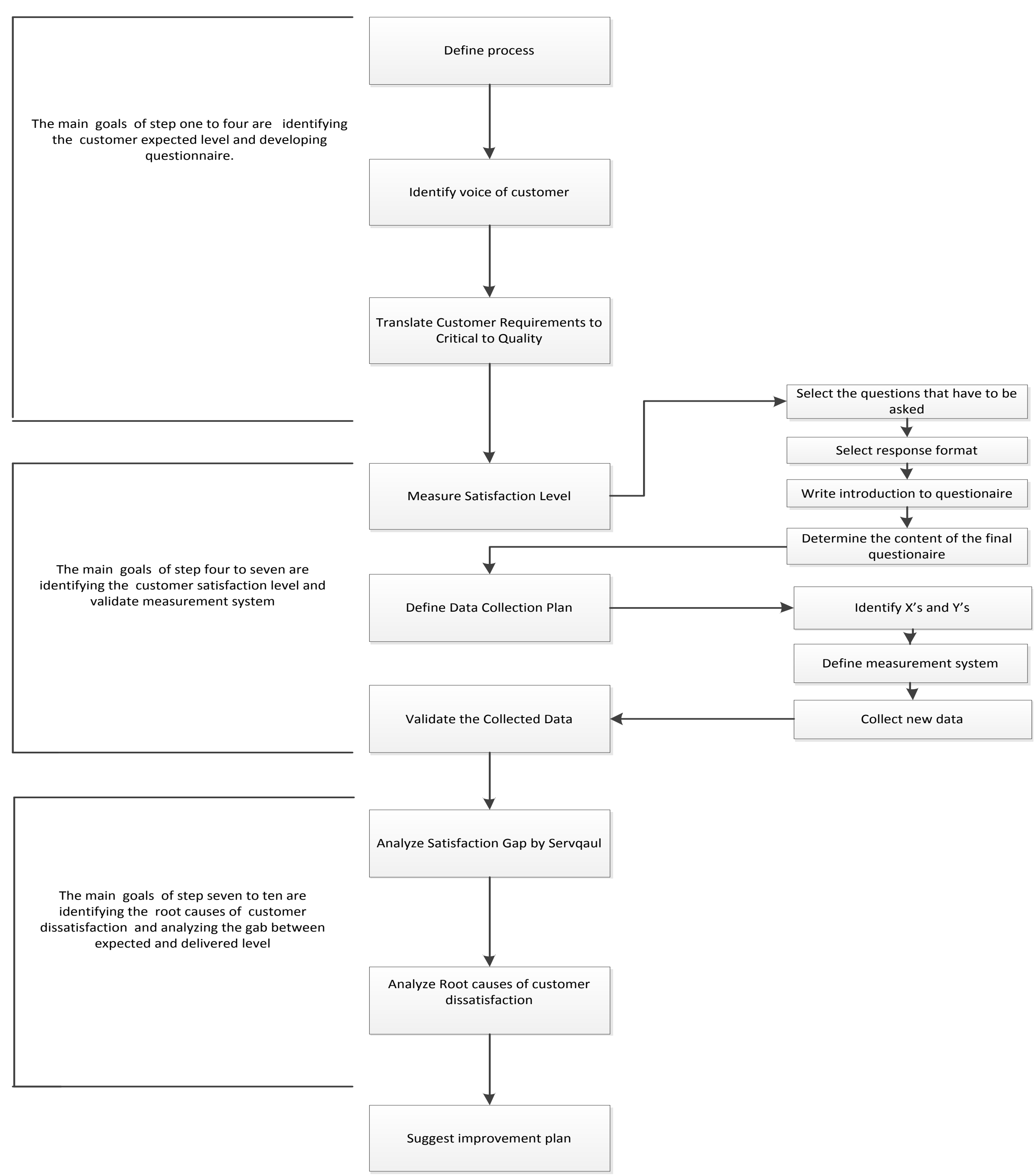

Figure 4.1 Summery of Framework Steps

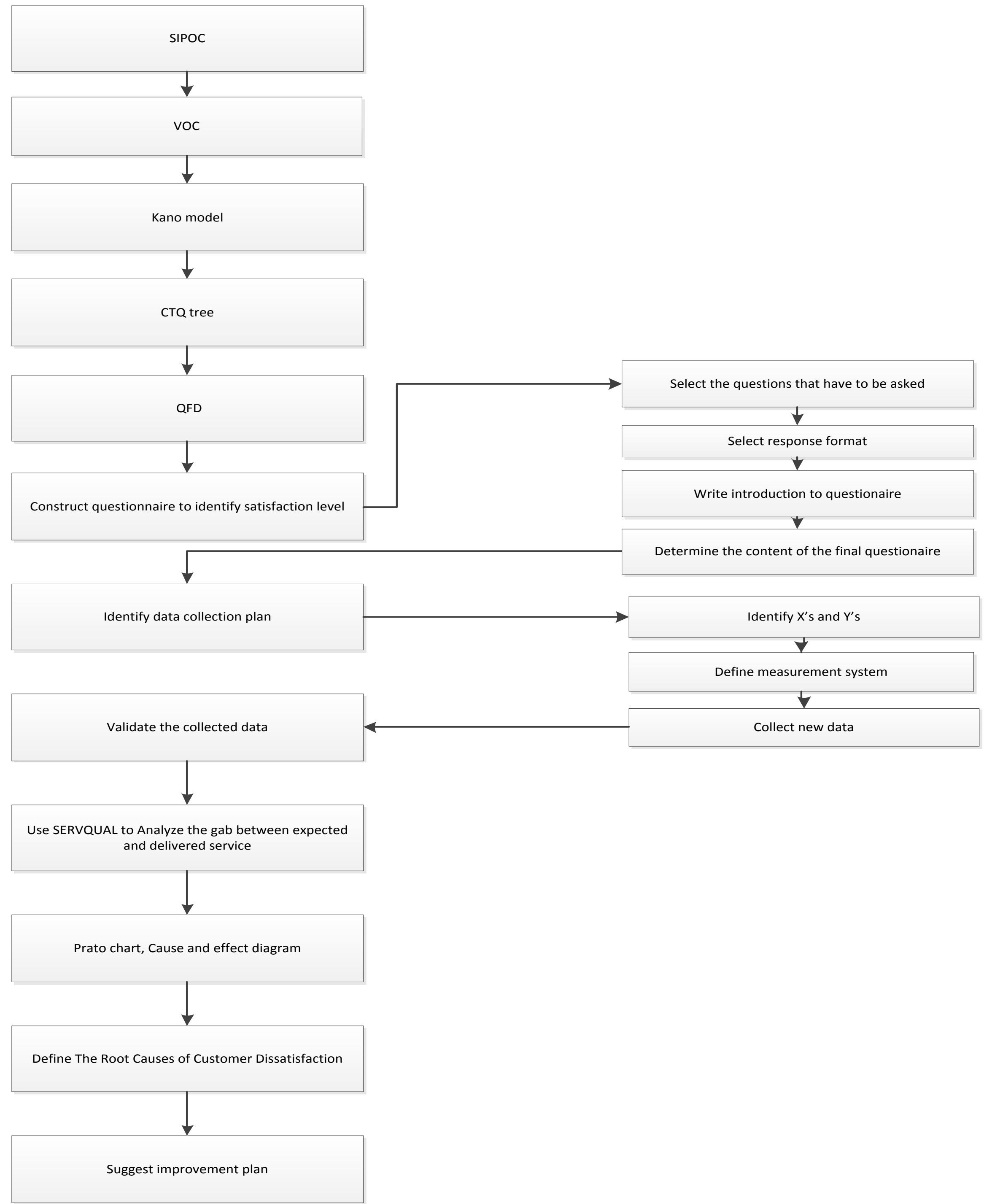

Figure 4.2 Framework's Quality Tools

### 4.3.1. Define the process variables:

This step is considered a critical and fundamental step to identify the major customers and understand their basic requirements and needs. Furthermore, the current process will be mapped and major stakeholders will be identified. In return, this will provide the system analyzers a good viewpoint about where the strengths and weaknesses are located in the system, in order to develop a good foundation for next steps. SIPOC has been used to define the process variables in computer maintenance services.

#### 4.3.1.1 SIPOC:

As mentioned before, the acronym SIPOC stands for Supply, Input, Process, Output and Customer which is used to provide a high level process map. The main reason for deploying SIPOC in this framework is to identify the customers who will be affected by the provided service. Also, it is a useful tool to identify major customer requirements and other relevant elements for the process. The table below illustrates the SIPOC diagram for most computer maintenance organization.

Table 4.6 SIPOC for Computer Maintenance

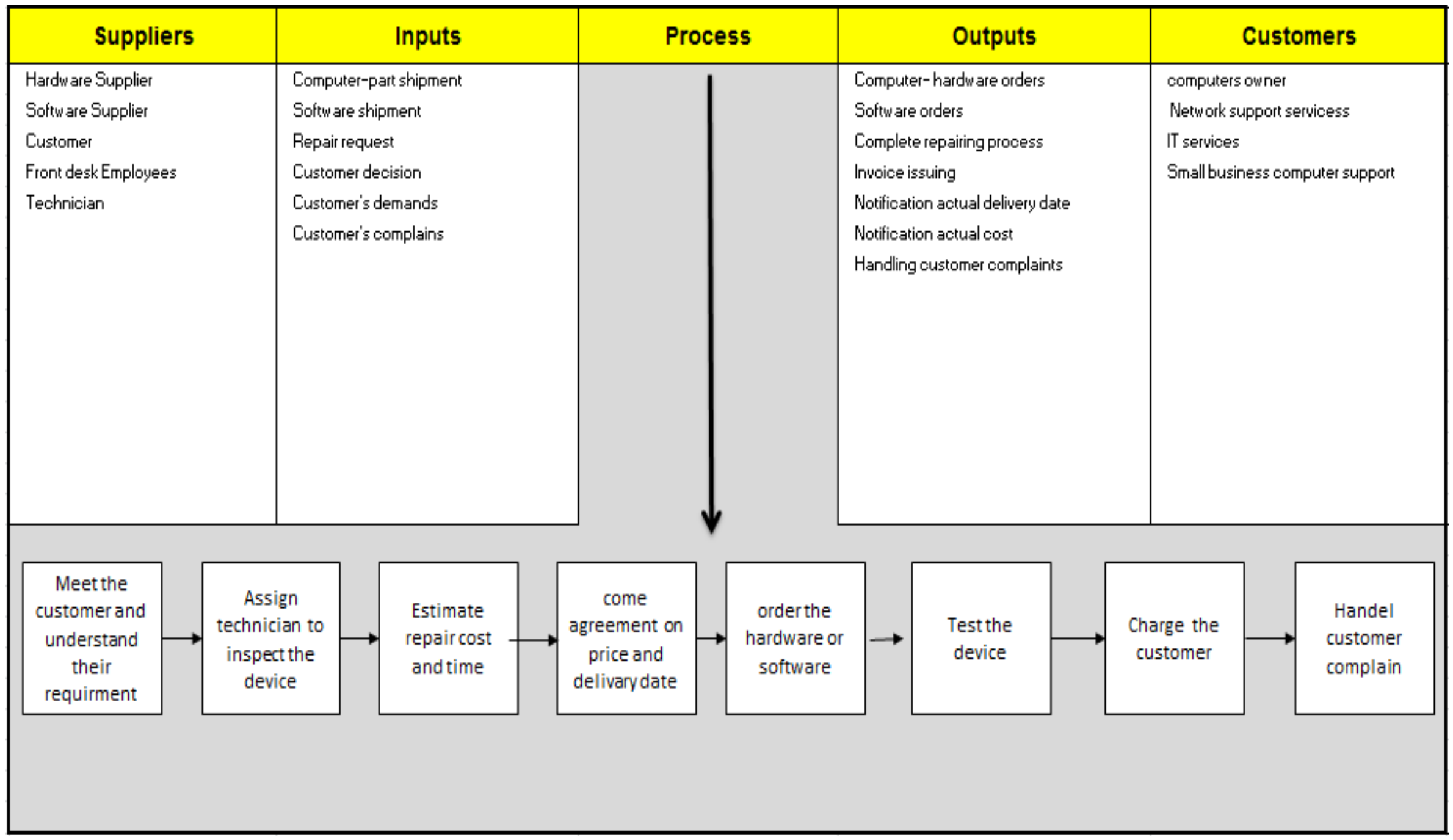

| Suppliers | Inputs | Process | Outputs | Customers |
|---|---|---|---|---|
| Hardware Supplier | Computer-part shipment | | Computer- hardware orders | computers owner |
| Software Supplier | Software shipment | | Software orders | Network support servicess |
| Customer | Repair request | | Complete repairing process | IT services |
| Front desk Employees | Customer decision | | Invoice issuing | Small business computer support |
| Technician | Customer's demands | | Notification actual delivery date | |
| | Customer's complains | | Notification actual cost | |
| | | | Handling customer complaints | |

Meet the customer and understand their requirment → Assign technician to inspect the device → Estimate repair cost and time → come agreement on price and delivary date → order the hardware or software → Test the device → Charge the customer → Handel customer complain

4.3.1.1.1 Supplier:

Two groups of suppliers to the process have been identified, the internal and external. Internal suppliers, refers to computer technicians and employees on the front line who deal directly with customers. External suppliers can be divided into three subgroups. The first group is the hardware providers which represent the organizations and retailers that sell computer components such as processors, hard disks, network adapters, etc. The software providers are considered another type of external supplier, representing all companies and stores that provide computer maintenance with security packages, operation systems, and other software programs. The customers demonstrate the third subgroup, which provides the two critical types of input, to be explained in next section.

4.3.1.1.2 Input:

The above supplier will provide the process with three different major groups of process inputs listed below:

1. Part Shipment: Typically, computer maintenance companies don't have all the hardware and software needed to complete the repairing process, so they often order from external organizations or shops. This reveals two types of process inputs which are time and cost.
2. Three types of input are connected to the customer. The first type of input is the customers' demand to receive their service right the first time. The second input is the customers' complaint which represents the customers' disappointment with the provided services either because the service is overcharged or the technicians did not do their job correctly. The third input is the customers' decision whether they are going to stay or go to another service provider.
3. The employees provide the process with four major inputs. The first input represents employees' professionalism in handling customer complaints. The second input represents the skill of the technicians for avoiding errors and delivering high quality services. The third input involves the employees' availability to assist the customer. Finally, the fourth major input is represented by response speed and the employees' attitude in answering customer inquiries.

#### 4.3.1.1.3 Process

By reviewing the process model for many computer maintenance companies six high-level steps have been highlighted:

1. Communicating with the customer and trying to understand their major requirements. The employees in the frontline should have excellent communication skills.
2. Assigning the case to a technician to diagnose and repair the problem in the device, the networks, or the database and conducting the inspection.
3. Informing the customer with an estimated cost and delivery date. If the price is too high or the delivery time too long, the customer may choose to leave. However, unrealistically promising the customer delivering their computers in a short amount of time for less money will cause dissatisfaction and cause customer decay.
4. Ordering the part from the suppliers. Choosing reliable suppliers is the essence of this step because this will impact customer satisfaction. Commitment, cost and consistency are critical criteria that should be considered when selecting the supplier.
5. Testing the devices or the services before they are delivered to the customer is a critical step to ensure the computer service is reliable and free of error.
6. Delivering the service at the promised time and ensuring the customer is satisfied is the last step. The company should use an appropriate payment method and ensure the customer payment information is secured.

4.3.1.1.4 Output:

Computer maintenance organization outputs can be classified into three major groups:

1. Customers' decisions whether to use the service or not represent. Many studies show that employees on the frontline have a significant impact on customer decisions. Those studies pinpoint two factors that affect customers' decisions to make purchases, which are employees' courtesy and employees' empathy. However, a considerable number of researchers argue that service cost is the main factor impressing the customers' decisions and it surpasses the previous two factors.
2. Ordering hardware and software from outside vendors. The major outputs of the ordering process are purchases' invoices, transactions and orders tracking.
3. The most critical output is providing the customer with reliable service; providing the customer with expected service at the promised time not only increases customer satisfaction, but also increases the organization profit by maintaining customer loyalty.
4. Gaining customer confidence is the final output. The customer should be informed exactly when the service is delivered. Moreover, all promises that have been made should be fulfilled. In addition, customers' complaints should be handled professionally.

4.3.1.1.5 Customer:

Commonly misunderstood, the customers are only those who buy products or services which reflect just external customers. The other type is internal customers which represent those who are directly connected to the organization. Internal customers in computer maintenance organizations are represented by front desk employees, technicians, engineers, and administrators. The external customers are represented by government agencies, local businesses and personal users. The maintenance process could be done on-site for the first three types of customers, so the company should have enough technicians to send out. Unlike in-shop services, those technicians will be directly in touch with the customers, they should be trained to be courteous.

### 4.3.2 Identify voice of the customer:

Voice of the customer can be defined as a way for acquiring the customer's needs, requirements and expectations. Capturing and translating voice of the customer is a substantial step to evaluating the process performances and identifying the root causes of customer dissatisfaction, which is the essence of services or products success. Furthermore, it guides the organization to identify what the customers' expectations are and what features are required the most. Voice of the customer research consists of four steps (see Figure 4.3). At the first step, all the internal and external customers should be pinpointed. Aforementioned, SIPOC is a most widely used tool for that purpose. Later on, the major customer requirements and needs for those customers should be identified accurately. Next, the importance of those needs should be assessed. The most popular methods of gathering data are Focus Group, surveys, and verbal communication. The last step is analyzing the collected data. The Kano model and quality function deployment is most commonly used tools to analyze the customer requirements.

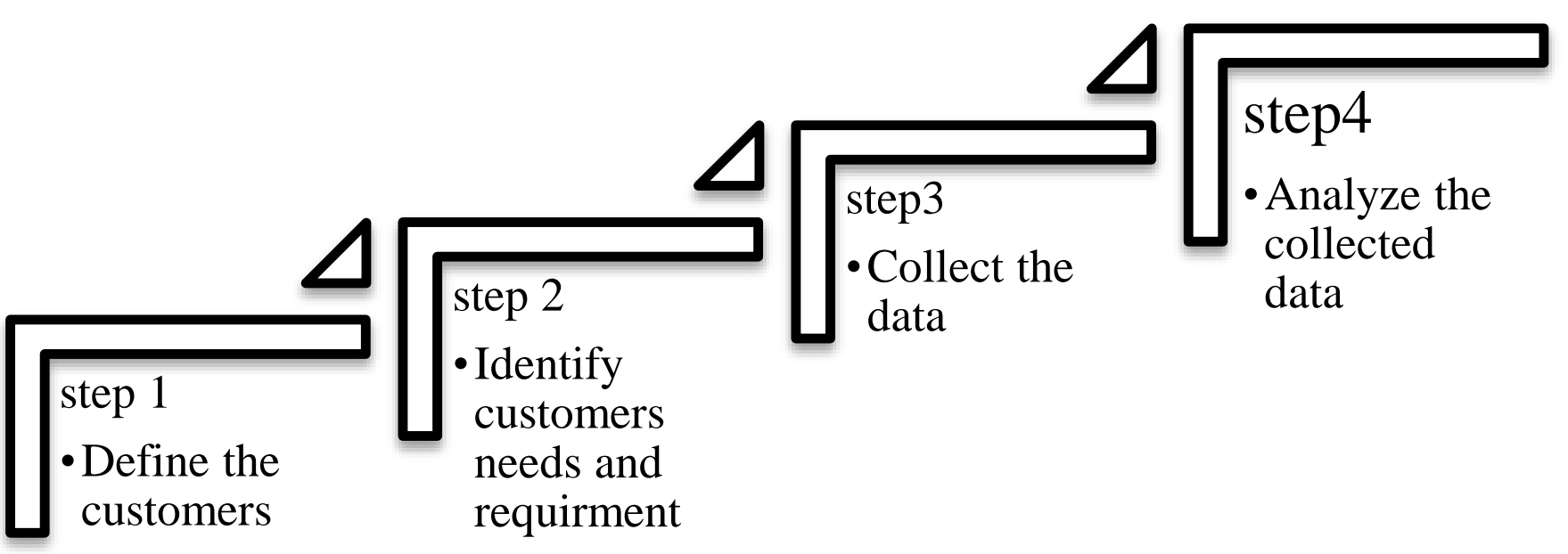

Figure 4.3 Voice of Customer Steps

Back to the framework, the key players have been already defined by using SIPOC, so the organization should know who the customers should care about. The next step is identifying the customer's requirement which is considered a critical step. At this step, the features of the services that the customers care about the most should be identified accurately by asking them what they expect. In other words, the relevant quality characteristics (the customer satisfaction key drivers) should be pinpointed. Previously, in section 4.2, five fundamental key characteristics in the computer maintenance industry have been identified which represent the major customer requirements and needs. Furthermore, those key drivers are subdivided in 28 items. Since those 28 items have been developed to evaluate different types of computer maintenance organizations, it is not necessary to ask the customer to evaluate all of them. Figure 4.4 illustrates the VOC items at computer service industry. Later on, the customer should be asked to evaluate each item on the 5 point Likert scale. The lowest rate is 1 which means strongly disagree and the highest 5 which means strongly agree. Other questionnaires should be developed to measure the importance and expected level for each key driver. Next, collecting the voice of customer should be started by carrying out the surveys that are developed in previous steps. Those data could be gathered in different ways such as email, social media, mail, telephone, or in-person. Many studies show that in-person and telephone are the finest ways to collect the voice of customer. After all, those methods are most expensive and limited to the existing or known customers. For that reason, many organizations prefer using email, online surveys and social media such as tweeter and Facebook to gather the voice of customer, but these methods have low quality responses, which explain why many organizations still use in-person and telephone call methods; however, the cost of conduct survey is the major factor that affects corporate management in selecting survey method. Afterward, the Kano model or QFD will be used to analyze those collected data.

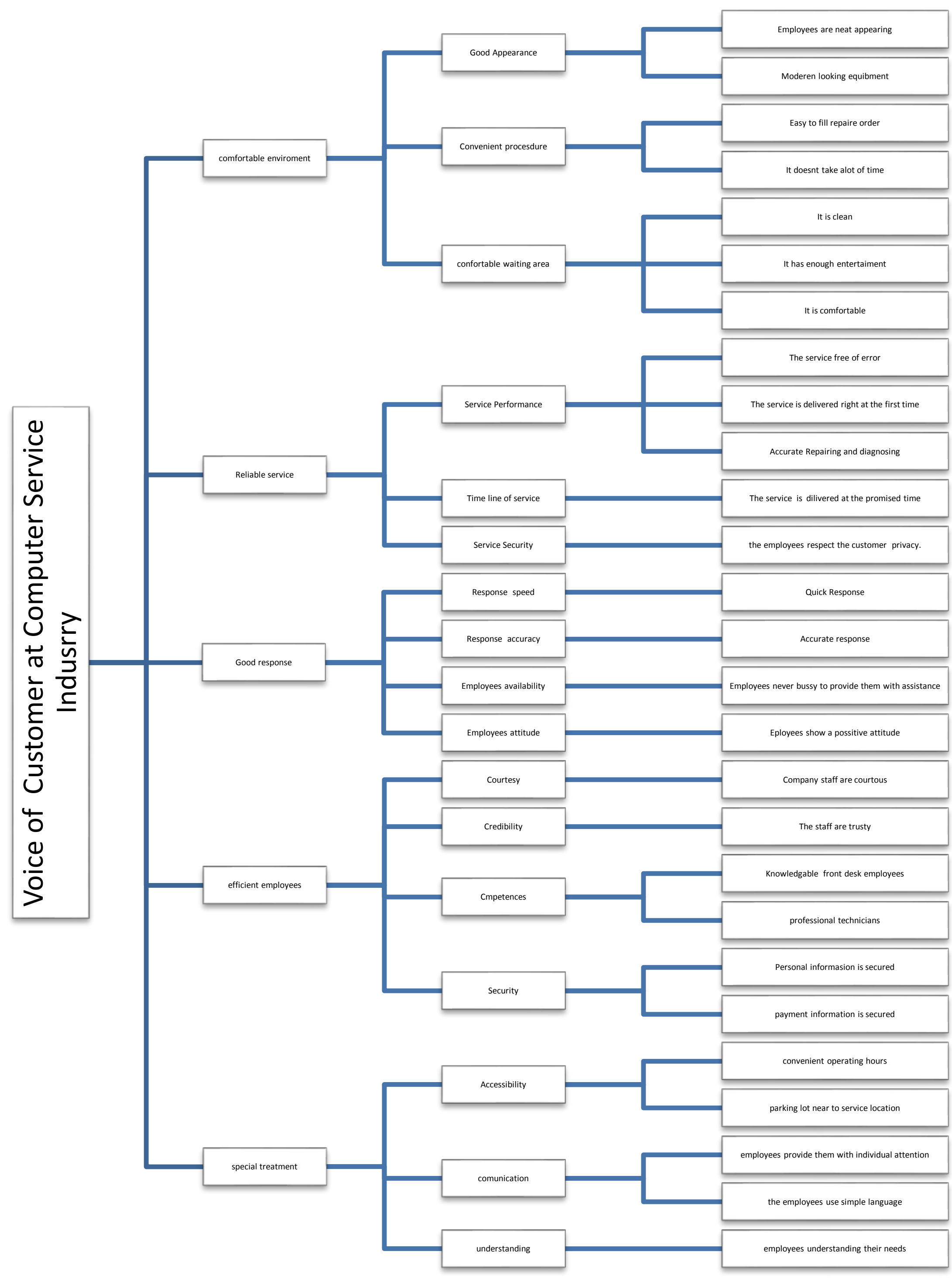

Figure 4.4 Voice of Customer at Computer Maintenance Organization

4.3.3. Categorize customer requirements by Kano model:

As we explained in literature review, Kano is a powerful tool to identify and analyze key characteristics of customer requirements and needs. The main goal of deploying the Kano model in this framework is to identify which service attributes can be used to achieve a high level of customer satisfaction. The Kano model will assist the computer maintenance organizations to distinguish among those requirements and needs. Based on the Kano model, the customer requirements at computer service organizations can be grouped into three major categories:

Expected Requirements are those which must be satisfied or included in the service. Satisfying those requirements will not increase customer satisfaction; however failing to deliver them will have a huge impact on the satisfaction level. Since satisfying those will not increase the satisfaction level, the organization should not increase a lot above the expected level. Reliability dimension is a good example for those requirements. For instance, the customers expect accurate repairing and diagnosing. If the technician does his job accurately that will not increase the satisfaction because the customers expect that in the service. On the other hand, failing to do those will make the customer extremely dissatisfied. The same is true for service security, employees' credibility, and technicians' competence. Table 4.7 shows a list of must be requirements in the computer service industry.

Revealed Requirements (performance requirements) are called One-dimensional requirements; in the computer service industry, those represent the spoken requirements. Customer satisfaction level will increase or decrease in proportion to fulfillment of those requirements. In other words, the higher level of fulfillment leads to a higher level of satisfaction. A good example is repair speed - the faster the service, the higher the customer satisfaction and vice versa. Another example is employees' courtesy - the more courtesy the better. Table 4.7 illustrates those requirements in the computer service industry.

Exciting Requirements represent unspoken requirements. Also, they are called delighters. Usually, the customers are unaware of those requirements, so failing to fulfill those requirements do not have any impact on the level of satisfaction; however, discovering and satisfying those requirements will extremely gratify the customer. Offering free services or availability of specific entertainment in waiting area is a good example for those requirements.

Table 4.7 Categorize Customer Needs by Kano Model

| | Characteristics | Kano attributes |
|---|---|---|
| Tangibles | Employees appearances | Delighter |
| | Visual aspect of Equipment | Delighter |
| | Difficulty to fill out the repair order | Must be |
| | Timely manner to fill order | Must be |
| | Cleanliness Level of waiting area | Must be |
| | entertainment in waiting area | Delighter |
| | Comfortable waiting area | Performance |
| Reliability | Error free service | Must be |
| | delivering service the service right at the first time | Must be |
| | Accuracy level of diagnosing and repairer | Performance |
| | Accuracy level of delivering the service | Must be |
| | The level of customer privacy | Must be |
| Respons e | Speed level of response | Performance |
| | Accuracy level of response | Must be |

| | Characteristics | Kano attributes |
|---|---|---|
| | Employees availability to assist the customer | Must be |
| | Employees attitude toward the customers | performances |
| Assurance | employees Courtesy | Performance |
| | Trusty Employees | Must be |
| | Knowledge employees | Performance |
| | Competence employees | Performance |
| | customer information are secure | Must be |
| | Payment information are secure | Must be |
| Empathy | convenient operating hours | Performance |
| | convenient service location | Must be |
| | personal attention | Delighter |
| | The difficulty of the language that is used in communication | Must be |
| | understanding customer needs | Performance |

### 4.3.4. Translate voice of customer into critical to quality:

In order to meet the customer requirements and develop competitive service, the firm has to use a structured approach to translate the voice of customer into technical requirements. Those requirements should prioritize and satisfy later according to the importance that customers give to them. This step has been designed to transfer the collected data from previous steps into performance drivers by using both CTQ tree and QFD.

#### 4.3.4.1Critical to quality tree:

CTQ tree is an effective tool that helps computer service organizations to deliver high quality service. It always starts with the customer requirements then moves forward to identify the performance requirements. The main goal of deployment CTQ tree is to find the technical requirements that should be available in computer services in order to rate them later by using QFD.

#### 4.3.4.2 Quality function deployment:

In the previous steps, the customers' requirements and their importance have been defined, and the customer-prioritized list of customer requirements is generated from the voice of the customer. Translating those requirements into quality characteristics is the next step. Quality function deployment (QFD) is the most rigorous and scientific method for translating the voice of the customer. Unlike the traditional quality system which attempts only to minimize the negative quality, QFD concentrates on maximizing the customer satisfaction. It assists the organizations to improve their services by searching on both spoken and unspoken needs and translate them into utilizable services (Mazur, 1993).

The main reason for deploying QFD in this framework is to translate the customer requirements (What the customer wants) into technical requirements (How to satisfy these requirements). In general, QFD consists of several steps. At first, the customer requirements and their importance should be identified. Section 4.2.2 of the framework has been designed to identify the customer requirements at computer maintenance organizations. The next step is specified to determine how each customer’s need will be satisfied. CTQ tree has been used to identify the technical requirements for each need. Developing a relationship matrix between customer requirements and technical requirements is the next step. In many cases, each of the customer requirements has more than one relationship with the technical requirements; that can cause confusion in determining the relationship between technical and customer requirements. Those relationships have been represented by using symbols whereas a black circle represents a strong relationship, a hollow circle represents a medium relationship, and a triangle represents a weak relationship. The next step is developing the correlation matrix which displays the relationship between the technical requirements. It is represented by the roof of QFD as shown in Figure 4.5. Also, symbols have been used to demonstrate the relationships between the technical requirements. Later on, both the customer and technical requirements are compared against the service competitors. The last step is determining the importance of weighing each technical requirement. At the end of this step, we transform the customer requirements from the previous step into a deliverable action.

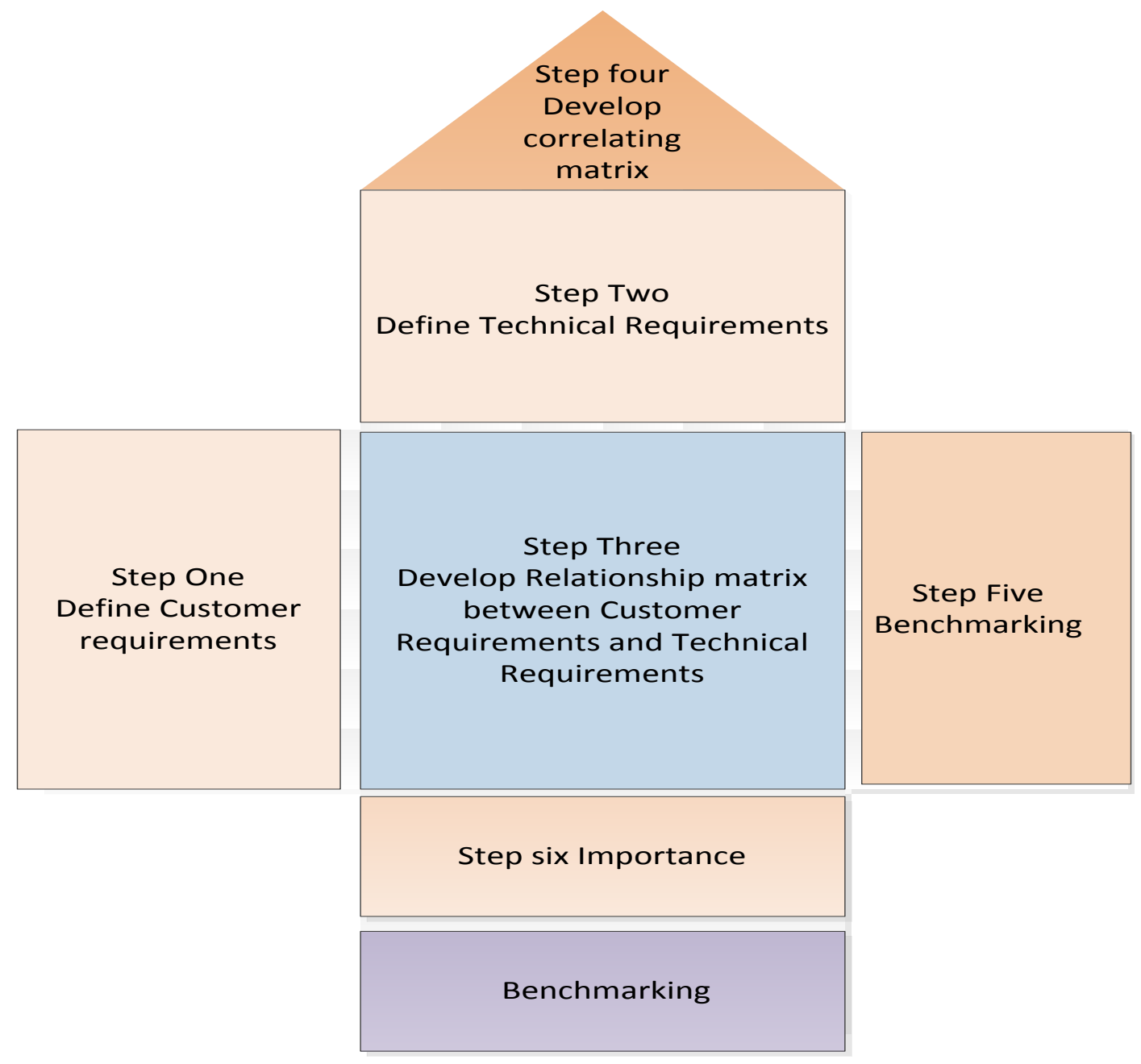

Figure 4.5 Quality Function Deployment Steps

## 4.3.5. Measure Perception level:

As mentioned earlier, the previous step is specified to capture and translate customer expectations and needs into technical requirements; however it does not assist the service providers to acquire and understand the customer's feeling in regards to their services. For that reason, this step has been developed to provide computer service companies a good insight about how the customers feel regarding the company's environment, employees, and provided services. In return, this assists the corporate managements to identify the strengths and weaknesses in the organization performances. Since measuring customer perception pinpoints the root causes of dissatisfaction, it is a fundamental step for avoiding the customer decay and keeping their existing customer satisfied. Furthermore, it assists the firm to evaluate its performance regarding customers' priorities.

Customer perception surveys are the most effective tool for acquiring the satisfaction level; however, the success of those surveys extremely depends on how well the questionnaires are designed. Hayes defines four basic phases to construct a successful customer perception questionnaire. At the beginning, the questions should be constructed from the customer's perspective of the service. Five critical areas should be evaluated at computers services; Tangibles, Reliability, Responsiveness, Assurance and Empathy. For that purpose, a 28 item questionnaire has been developed in the preceding section (see section 4.2). The following criteria should take into consideration when customer perception surveys are constructed. First of all, the questions should be comprehensible. A simple language and specific statement in the questionnaire has been used to avoid ambiguity. Furthermore, each question should be used to evaluate no more than one key characteristic (customer requirement), so it is incorrect to ask the customer to evaluate two or more characteristics at the same time. For example, did the employees answer your inquiry quickly and accurately? The previous question is incorrect because the customer was asked to evaluate two different requirements (employees' response and employees' competence) at the same time; for that, it must be split into two questions. Lastly, the questionnaire should be concise whereas a long wordy question makes the survey long and inconvenient; in return, it contributes low response rate. Selecting the response format is the next phase. The response format refers to the way the data is collected from the questionnaires. There are many scaling methods; Likert, Guttman, and Thurstone methods. Since Likert method provides the customer with the freedom to evaluate the service in varying degrees, it is the recommended method to evaluate the service at computer maintenance organizations. Writing introduction to questionnaires is the third phase. It is specified to explain the main purpose of the questionnaires and provide the customer with instructions to complete them; also, it has to be simple and short as well. The last phase is item selection. Usually, many questions are generated to evaluate the customer satisfaction. For instance, the author identified 28 items in the computer service industry (see section 4.2). It would be hard to make the customers answer all of these questions. Selecting the questions should be based on the overall representation for customer requirements. In other words, the question that represents the customer requirements the most should be included.

#### 4.3.6. Define data collection system:

Unlike manufacturing processes, the service processes completely depend on the interaction between the employees and customers. It is easier to collect the data like setup time than to capture satisfaction level. Mostly, it is hard to make the customers to complete the survey which is the main source of customer data. In many cases, the computer service companies depend on the customer complaints as a main source of data; however it just reflects the negative feedback of customers that illustrates only the dark side of service. Also many studies show that dissatisfied customers are more motivated to complete surveys than satisfied customers.

Unless the organizations motivate their customer to complete the surveys, those surveys are unreliable because they don't reflect the opinion of the majority. Many tactics have been developed to motivate the customers to take the surveys. One way to motivate the customers is by displaying the interest in their opinion and revealing the importance of their participation for service improvements. Give them a reason to complete the surveys, which should be simple and short. Another way to motivate the customer is by offering incentives or rewards.

There are many ways to encourage customers to take surveys, in-person, by mail, or email for example. The first way contributes a high number of respondents, but is limited to specifically in-shop customers. In-person surveys are usually initiated at the front desk when an employee asks a customer to complete the survey while they wait, or mail it later. The company may also mail or email the customer asking for feedback.

### 4.3.6.1 Define Measurement System

At the previous steps, Performance Metrics have been identified. The X's represent the independent variables that affect the satisfaction level. We might recall from previous sections that key characteristics of customer satisfaction depict X's. On the other hand, the Y's are dependent variables which are driven by X's. By way of explanation, the overall customer satisfaction in computer maintenance(Y) depends on the X's. In other words, Y is function for x "$Y = f(x)$". As previously illustrated, the satisfaction level can be defined as the difference between perceived and expected value -the higher the positive value, the higher the customer's satisfaction and vice versa. In other words, the satisfaction level for each item can be found by subtraction of the perceived level from the expected level that is to say "$Y_i=X_{in} - X_{ni}$." Table 4.8 illustrates the customer satisfaction measurements of the computer service organization.

Table 4.8 Independent Variables (X's)

| Dimension | Perception | | Expectation | |
|---|---|---|---|---|
| | $X_{ni}$ | Description | $X_{in}$ | Description |
| **Tangibles** | $X_{n1}$ | Employees appearances | $X_{1n}$ | Employees appearances |
| | $X_{n2}$ | Visual aspect of Equipment | $X_{2n}$ | Visual aspect of Equipment |
| | $X_{n3}$ | Difficulty level to fill repair order | $X_{3n}$ | expected level to fill repair order |
| | $X_{n4}$ | Timely manner to fill order | $X_{4n}$ | Timely manner to fill order |
| | $X_{n5}$ | Cleanliness Level of waiting area | $X_{5n}$ | Expected level of waiting area Cleanliness |
| | $X_{n6}$ | entertainment level in waiting area | $X_{6n}$ | Expected level of entertainment in waiting area |
| | $X_{n7}$ | Comfort level of waiting room | $X_{7n}$ | Expected level of Comfort waiting room |
| **Reliability** | $X_{n8}$ | Error free service | $X_{8n}$ | Error free service |
| | $X_{n9}$ | Accuracy level of delivering service the service right at the first time | $X_{9n}$ | Expected accuracy level of delivering service the service right at the first time |
| | $X_{n10}$ | Accuracy level of diagnosing and repairer | $X_{10n}$ | Expected accuracy level of diagnosing and repairer |
| | $X_{n11}$ | Accuracy level of delivering the service | $X_{11n}$ | Accuracy level of delivering the service |
| | $X_{n12}$ | The level of customer privacy | $X_{12n}$ | Expected level of customer privacy |
| **Response** | $X_{n13}$ | Speed level of response | $X_{13n}$ | Expected speed level of response |
| | $X_{n14}$ | Accuracy level of response | $X_{14n}$ | Expected accuracy level of response |
| | $X_{n15}$ | Employees availability to assist | $X_{15n}$ | Employees availability to assist the customer |

| Dimension | Perception | | Expectation | |
|---|---|---|---|---|
| | | the customer | | |
| | $X_{n16}$ | Employees attitude toward the customers | $X_{16n}$ | Employees attitude toward the customers |
| Assurance | $X_{n17}$ | The level of employees Courtesy | $X_{17n}$ | Expected level of employees courtesy |
| | $X_{n18}$ | The level of employees credibility | $X_{18n}$ | Expected level of employees credibility |
| | $X_{n19}$ | The level of employees knowledge | $X_{19n}$ | Expected level of employees knowledge |
| | $X_{n20}$ | Competence level of employees | $X_{n20}$ | Expected Competence level of employees |
| | $X_{n21}$ | Safety level of customer personal information | $X_{n21}$ | Expected Safety level of customer personal information |
| | $X_{n22}$ | Safety level of customer payment information | $X_{n22}$ | Expected Safety level of payment information |
| Empathy | $X_{n23}$ | The level of convenience of operating hours | $X_{n23}$ | Expected level of convenience of operating hours |
| | $X_{n24}$ | The level of convenient of service location | $X_{n24}$ | Expected level of convenient of service location |
| | $X_{n25}$ | The level personal attention | $X_{n25}$ | Expected level personal attention |
| | $X_{n26}$ | The difficulty of the language that is used in communication | $X_{n26}$ | Expected level difficulty of the language that is used in communication |
| | $X_{n27}$ | The level of understanding customer needs | $X_{n27}$ | Epected level of understanding customer needs |

### 4.3.7 Validate the collected data

This step is specified to ensure the collected data from previous steps are accurate, precise, repeatable and reproducible. Unlike the physical measurement instruments, by nature customer satisfaction surveys have a sort of variability or bias in their measurements due to either the customer misinterpretation of rating scale or bias in customers' attitude to evaluate the service. For instance, the customers are more motivated to participate in the surveys if they receive either bad or extremely good service than the people who receive regular service. Also, some customers tend to be more positive or negative than others, so the collected data from customer satisfaction surveys should be tested before they are implemented in the study. There are many methods that may be used to validate the customer satisfaction surveys such as Gage R&R, Hypothesis testing, etc. The following hypothesis could be used testing customer surveys, however, the cronboach alpha has been selected to validate customer satisfaction surveys.

$H_{0:}$ The customer surveys reflect the actual of customer perception.

$H_1$: The customer surveys data doesn't reflect the majority of customer perception.

### 4.3.8. Analyze the Gap by SERVQUAL

In the previous steps, performance metrics have been defined, and the data has been collected. Also, those data have been validated. This step has been designed to find the gap between perceived and expected levels. In 1988, Zeithaml, Parasuraman, and Berry developed a technique for evaluating the provided service in service organizations; since then this tool has been widely used by many service organizations. This tool has been selected to identify the gap in computer service organizations. In general, there are four steps to measure the customer satisfaction with the computer services. At the beginning, yi for each item should be calculated by subtracting the perceived level from the expected level ($Yi = X_{in} - X_{ni}$). Later, the average score for each dimension should be calculated ($Y_d$= average score for each dimension). In the constructed survey, the customers have been asked to evaluate the importance of each dimension ($I_d$). The weighted score ($W_d$) for each dimension will be found by multiplying the importance level by the average score, which was calculated in step two. At the end, the SERVQUAL score can be found by taking the average score for the five dimensions. The figure bellow illustrates the Gap Analysis framework. It also shows the performance metrics (Y's) for each step.

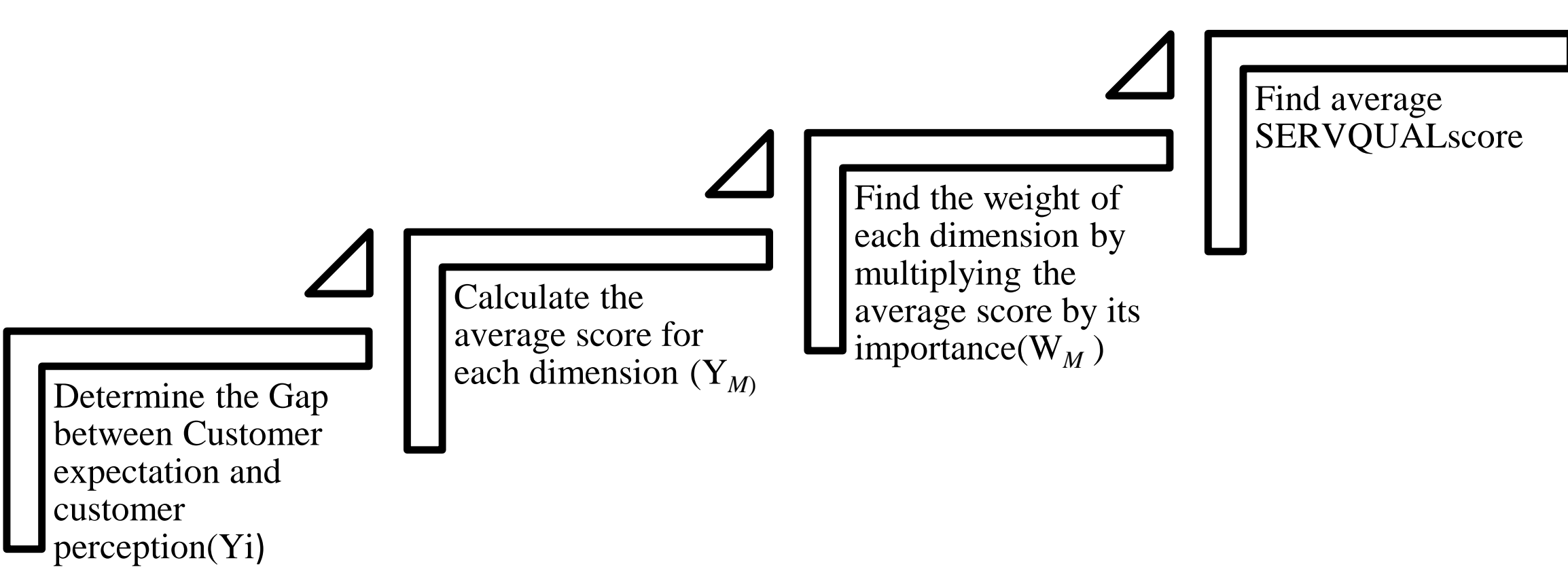

Figure 4.6 SERVQUAL Steps

### 4.3.9. Analyze the root cause of customer dissatisfaction

In the preceding step, the gap between customer perception and customer expectation for each customer requirements has been identified. The negative values mean the service provider has failed to fulfill the customers' expectation; in other words, they reflect customers' dissatisfaction with the service. This step has been developed to identify the root cause of customer dissatisfaction in computer service organizations. There are many factors that cause the organizations to fail to maintain their customers' loyalty. Cause and effect diagram is an efficient tool to uncover the main causes of customer dissatisfaction. However, it doesn't show the impacts of those causes. Furthermore, the impact for those causes is varied. For instance, the impact service reliability is much higher than the service tangibles. Usually, the companies satisfy or dissatisfy the customers' requirements at different levels. This reveals the needs to use another tool to analyze the dissatisfaction level. Pareto chart has been selected for this purpose. At the end of this step, the company will gain full understanding about the main causes of customer dissatisfaction with the service.  In addition, the strengths and weaknesses in each service area will be pinpointed. By identifying the weaknesses in the service, the organizations will be able to develop an improvement plan to optimize their performances.

# CHAPTER 5
# CASE STUDY

## 5.1 XYZ Company profile

XYZ Company is a small size company which was established in 1997. It is located in Baghdad. This company offers a variety of computer services which include both consulting and technical support that encompass computer diagnostics, computer support, computer repair, computer customization, computer clean-up, Operating System Re-install, hardware configuration, software installation, network installation, network repair and maintenance and computer business solutions. XYZ's customers can be categorized into two groups. The first group is personal computer users who represent the majority of the company's customers. The second type of customer is represented by local business and government agencies. The company offers both in-shop services for the first group and on-site services for the second group.

## 5.2. Problem statement:

XYZ severely suffers from both global and local competition. The administrators have noticed that most of their customers do not return. Furthermore, many employees claim that former customers have been absorbed by local competitors. However, a number of the managers argued they haven't received a large number of complaints regarding their provided service. Most of the managers agree the majority of customers are not loyal to their company, and the level of customer satisfaction is not compatible with the company expectation. At the same time, all the employees confirm that customer satisfaction is the key for long term success.

The previous section highlights a fundamental problem in many computer service companies. Many of XYZ Company administrators realize the importance of customer satisfaction for long-term success; however, they don't know what the main reasons for customer dissatisfaction are. This illustrates that the company is lacking a robust measurement system to gauge customer satisfaction level. Measuring the customer satisfaction is not an easy process. But, if XYZ Company identified the main causes for customer dissatisfaction, it would assist the company in improving their performance and boosting profit margin.

## 5.3. The Goals of this study can be summarized in the following points:

1. Identify both the expected and perceived customer satisfaction level in XYZ Company.

2. Identify the gap between customer perception and customer requirements at XYZ Company.

3. Identify the root causes of customer dissatisfaction at XYZ Company.

4. Suggest an improvement plan to enhance the company performance.

## 5.4 Framework Deployment:

To pinpoint the root causes of customer dissatisfaction and measure the level of satisfaction, the author has deployed the framework which has been developed in chapter four. The main purpose of deploying the proposed framework is to aid XYZ Company to improve the weakness area in company performance by highlighting and analyzing the main sources of customer dissatisfaction and interpreting the results into a robust improvement plan. Eight steps have been developed for that particular reason starting by identifying the major customers and ending with an improvement plan.

## 5.4.1 Define XYZ process variables

This step has been used to identify the process key players at XYZ Company. For that particular reason, SIPOC has been deployed to summarize the process input and output variables. Figure 5.1 illustrates SIPOC diagram for XYZ Company. The process has been mapped to illustrate interaction between the company departments and the customers.

### 5.4.1.1 SIPOC

Two groups of suppliers have been identified at XYZ Company. The first group is the internal supplier which is represented by the company employees. Furthermore, this group is divided into three subgroups which are the front desk employees, the technical support department, and the financial department. The external supplier is the second group which consists of both customers and hardware and software suppliers. Al-Nabaa, Aghadeer, SSC, Al-ajial, Al-muktar, Anas are the major hardware and software suppliers for XYZ Company.

Likewise, those suppliers provide the company process with three major inputs. The first category is associated with the internal suppliers, which include employees' courtesy, employees' availability, employees' attitude, employees' knowledge, technician professionalism and response speed. The second category is related to external suppliers that encompass the customers' requirements and needs, customers' complains, hardware and software delivery speed, and shipments information.

Seven high-level steps for XYZ process have been highlighted as has been illustrated in Figure 5.1. This process has 10 major outputs, which are customer decision, paper work, repair cost estimation, work order, quality of repair, delivery time, invoices, payment receipts, customer satisfaction, and customer feedback. The company has different types of customers, which are listed in two categories: internal and external customers. As previously described, the internal customers represent the company employees that include the front desk employees, technicians, engineers and even the company administrators. The second group is represented by government agencies, local businesses and individual users who use the company service and pay company bills; they are not part of XYZ Company.

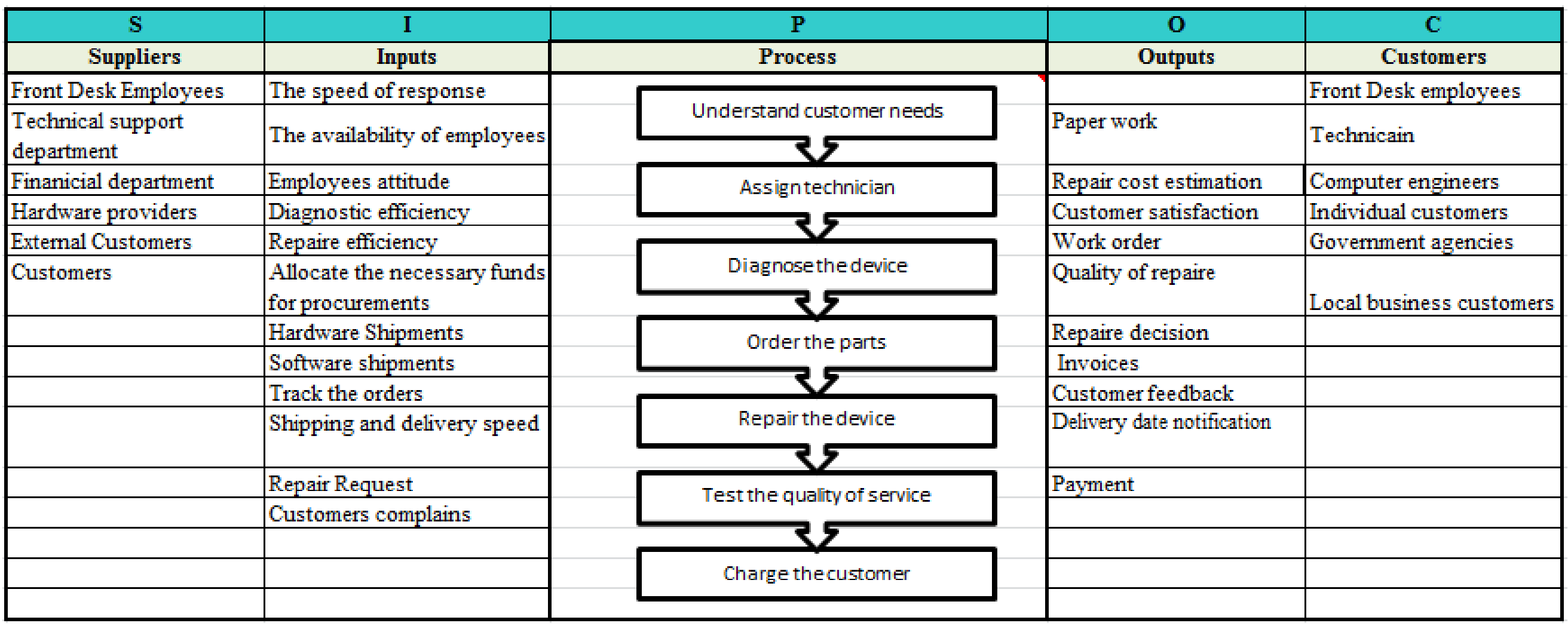

| S | I | P | O | C |
|---|---|---|---|---|
| **Suppliers** | **Inputs** | **Process** | **Outputs** | **Customers** |
| Front Desk Employees | The speed of response | Understand customer needs | | Front Desk employees |
| Technical support department | The availability of employees | | Paper work | Technicain |
| Finanicial department | Employees attitude | Assign technician | Repair cost estimation | Computer engineers |
| Hardware providers | Diagnostic efficiency | | Customer satisfaction | Individual customers |
| External Customers | Repaire efficiency | | Work order | Government agencies |
| Customers | Allocate the necessary funds for procurements | Diagnose the device | Quality of repaire | Local business customers |
| | Hardware Shipments | Order the parts | Repaire decision | |
| | Software shipments | | Invoices | |
| | Track the orders | | Customer feedback | |
| | Shipping and delivery speed | Repair the device | Delivery date notification | |
| | Repair Request | Test the quality of service | Payment | |
| | Customers complains | | | |
| | | | | |
| | | Charge the customer | | |
| | | | | |

Figure 5.1 XYZ Company SIPOC Diagram

5.4.1.2. Process map

To understand XYZ's process and to gain insight about how the customers interact with the company departments, the author has developed a process map for the company as illustrated in Figure 5.2.

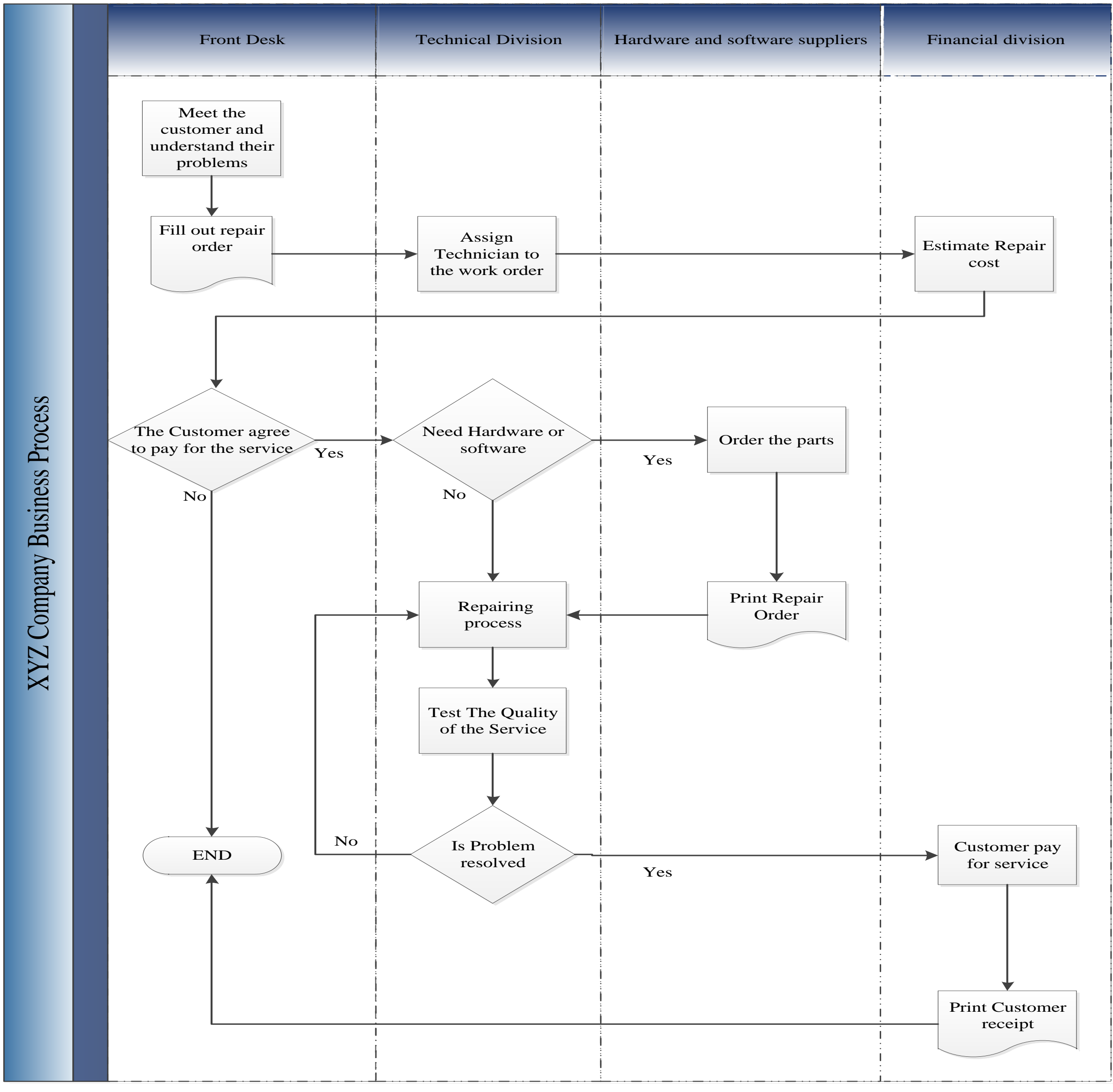

Figure 5.2 XYZ Company Process Flow

### 5.4.2 Identify expected level at XYZ Company

The main purpose of this step is to identify the major customers' requirements and their needs in XYZ Company. For that purpose, the author has used the steps that have been developed in a previous chapter to identify the voice of customer. At the preceding step the major company customers have been highlighted by using a SIPOC map. After the author gained a full understanding of whom the customers are and how they interact with the company, a customer expectation survey was constructed. This survey consists of two groups of questionnaires. The first group aims to determine the expected level in the service where the customer has been asked to evaluate 17 items that represent the relevant quality characteristics (see Appendix B). The second group is designed to capture the level of importance for each of the five dimensions.

Two types of rating scales have been used to evaluate those surveys. A five-point Likert scale ranging from 1 (extremely unimportant) to 5 (extremely important) has been used to evaluate the first group of questions. A different scale type has been employed to measure the importance score for each dimension – whereas 100 points have been allocated for the five dimensions of service quality and the customers have been asked to evaluate them according to their importance (see Appendix B). To increase the number of respondents and to make the survey more comfortable, the customer has been asked to use the multiples of five to simplify the mathematical operation.

Next, those surveys were handed manually to customers and they were asked to complete them manually (paper-based). The customer expectation surveys were distributed between October 10, through November 18, 2013 – 81 observations were collected. Subsequently, those observations have been saved on an Excel spreadsheet (see Appendix E). Minitab, Excel and SPSS software have been employed to analyze the collected data.

The first group is comprised of the first 17 questions. Those questions reflect the XYZ Company's necessity to highlight the level of expectation and understand the customer demands. Figure 5.3 and Table 5.1 illustrate the average score for each of these questions (or Items). Item one "deliver the services right at the first time" got the highest rating score which means this item is more difficult to satisfy than others. Figure 5.3 shows that both item 11 and 15 got the lowest score. Since customer expectation is low for the two previous items, then satisfying them will not be challenging. Based on these results, the company should pay more attention to the items with a high expectation level or a high rating.

Table 5.1 Customer Expectation

| **Item number** | **Descriptions** | **Average** | **Variance** |
|---|---|---|---|
| **1** | Accuracy level of delivering service the service right at the first time | 4.395061728 | 0.263679317 |
| **2** | Accuracy level of delivering the service at the promised time | 4.296296296 | 0.356652949 |
| **3** | Employees attitude toward the customers | 4.074074074 | 0.537722908 |
| **4** | Speed level of response | 4.345679012 | 0.349641823 |

| Item number | Descriptions | Average | Variance |
| --- | --- | --- | --- |
| **5** | Employees availability to assist the customer | 4.234567901 | 0.500533455 |
| **6** | Safety level of customer information | 4.185185185 | 0.471879287 |
| **7** | Reasonable repair cost | 4.197530864 | 0.55357415 |
| **8** | The level of employees Courtesy | 4.086419753 | 0.375247676 |
| **9** | The level of employees knowledge | 3.765432099 | 0.377076665 |
| **10** | The level of convenience of operating hours | 3.839506173 | 0.554488645 |
| **11** | The level of convenience of service location | 3.432098765 | 0.319463496 |
| **12** | The level personal attention | 3.481481481 | 0.545953361 |
| **13** | The simplicity of the language that is used in communication | 3.530864198 | 0.446578266 |
| **14** | The level of understanding customer needs | 3.728395062 | 0.370675202 |
| **15** | Employees appearances | 3.432098765 | 0.344154854 |
| **16** | Comfort level of waiting room | 4.172839506 | 0.365188234 |
| **17** | Visual aspect of Equipment | 3.580246914 | 0.367017223 |

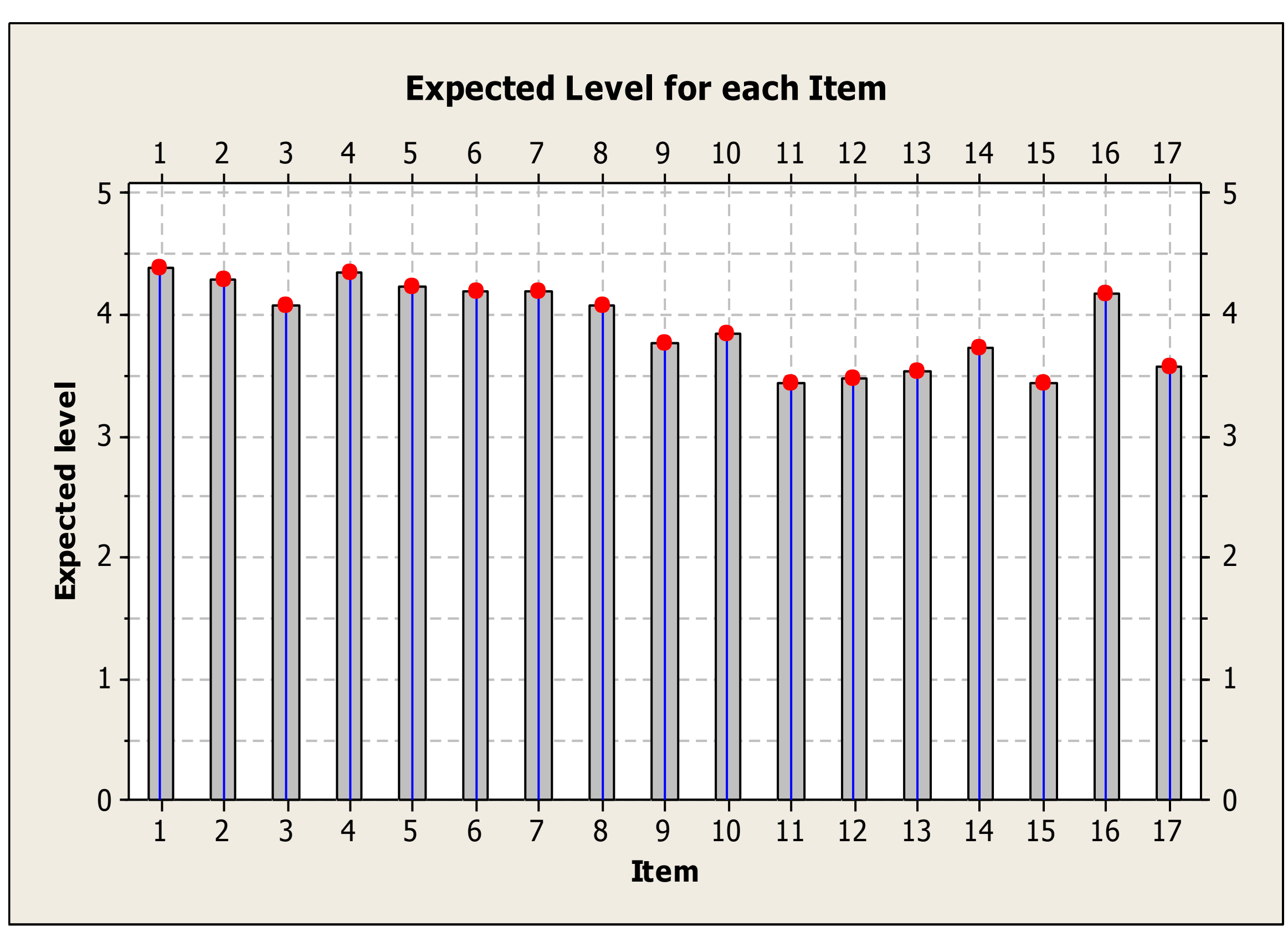

Figure 5.3 Expected Level for Each Item

Those 17 items are categorized into five groups (See Figure 5.4). The reliability dimension is represented by items one and two. To determine the expected level of reliability dimension, the average score for both items has been calculated. The second group encompasses question three through five which represent the service responsiveness. Based on table 5.2, the responsiveness and reliability dimensions have received the highest rating scores. Assurance dimension consists of question six through nine. Even though it ranked the third dimension, it got 4.05, which means the level of expectation is still very high. Empathy dimension received the lowest score among the five dimensions. It is comprised of question seven through fourteen. The last group is represented by tangibles dimension, which encompass questions 15 through 17.

Table 5.2 Expected Level for Each Dimension

| Dimensions | Average | Variance |
|---|---|---|
| **Reliability** | 4.345679012 | 0.191936728 |
| **Responsiveness** | 4.218106996 | 0.224537037 |
| **Assurance** | 4.058641975 | 0.107204861 |
| **Empathy** | 3.602469136 | 0.134166667 |
| **Tangibles** | 3.728395062 | 0.137838649 |

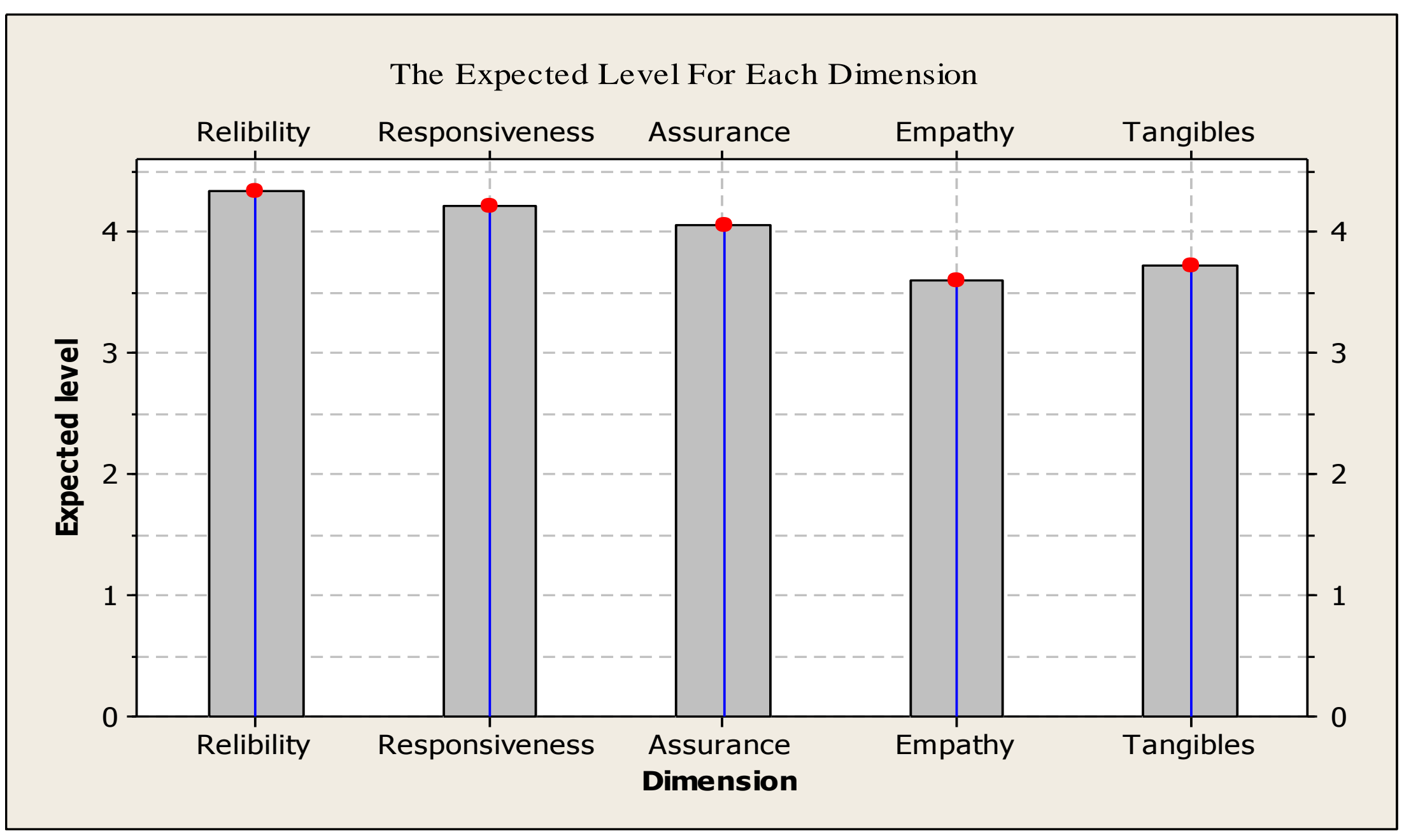

Figure 5.4 Expected Level for Each Dimension.

Based on table 5.2, and Figure 5.4, the average dimension score range from 3.60 to 4.3. However the importance (weight) of these dimensions is radically varied. For instance, the importance of service reliability is much higher than the company physical appearance. Since the impact of those dimensions on customer satisfaction is not equal, measuring expected level alone is not enough. For that reason, the second group of questions (Q17 to Q22) has been constructed to assess the weight of each dimension.

By way of explanation, the customer has been asked to rate the importance of each dimension. These results are illustrated in the Figure 5.5. The reliability dimension with a mean score of 39.69% was ranked the most important dimension, followed by responsiveness dimension which has an average importance score of 22.19%. In other words, those dimensions have the highest impact on customer satisfaction than the other dimensions.

The tangibles dimension has been ranked the least important dimension with an average score of 8.78%, meaning the influence of this dimension has the least impact on overall satisfaction. The main goal of measuring the relative importance of dimensions is to assist the company to determine the weighted gap score by multiplying them by the gap in service (Perception-Expectation). In return, this provides the company with proper insight about how each of those dimensions affect customer satisfaction.

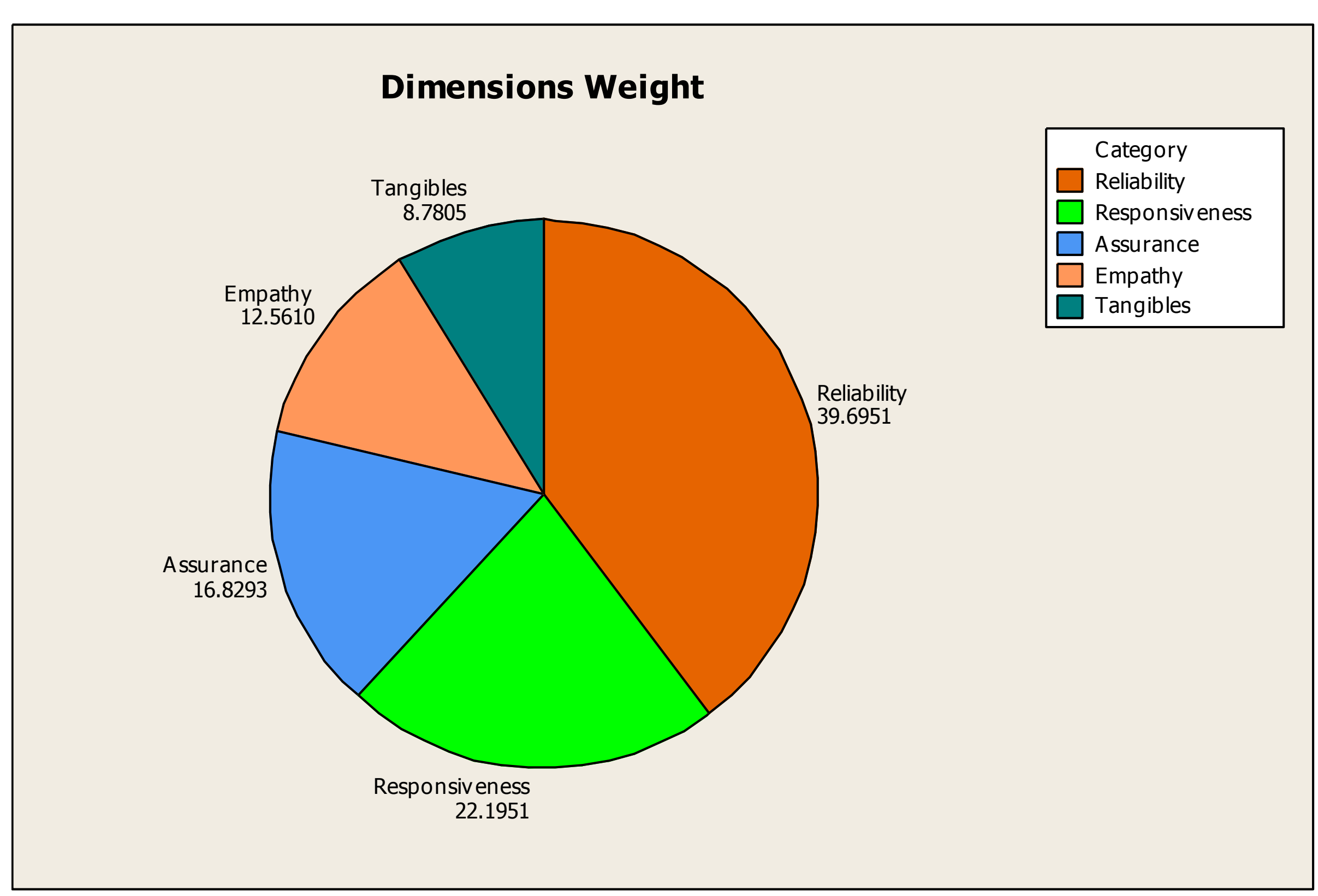

Figure 5.5 Dimension Weight

5.4.3 Translate voice of customer into critical to quality:

In order to develop the quality of provided service, QFD has been deployed to translate XYZ Company customers' requirements into technical requirements. The author has used the five steps to deploy house of quality in the service industry which has been described in detail in chapter four. Starting with the customer requirements, the author categorizes them into five groups labeled: tangibles, reliability, responsiveness, assurance and empathy. Then, the customers are asked to rank those dimensions according to their importance. Determining the service requirements that satisfy customer requirements was the next step.

Twenty characteristics have been pinpointed – as illustrated in Figure 5.6. Each of those requirements has a relationship with one or more customer requirements. To grasp how customer requirements and service requirements interact with each other, the relationship matrix has been created. Since the relations between customer and service requirements are not a one-to-one relationship, a four-point scale has been used to define the strength level. The scale varies from zero (no relation) to nine (strong relationship) between the technical and customer requirements.

The roof of the house of quality represents the correlation among the technical requirements. To evaluate the correlation between those technical requirements, a three-point scale has been deployed whereas the positive and negative signs have been used to define the correlation of the technical attributes. Figure 5.6 illustrates the three types of correlations.

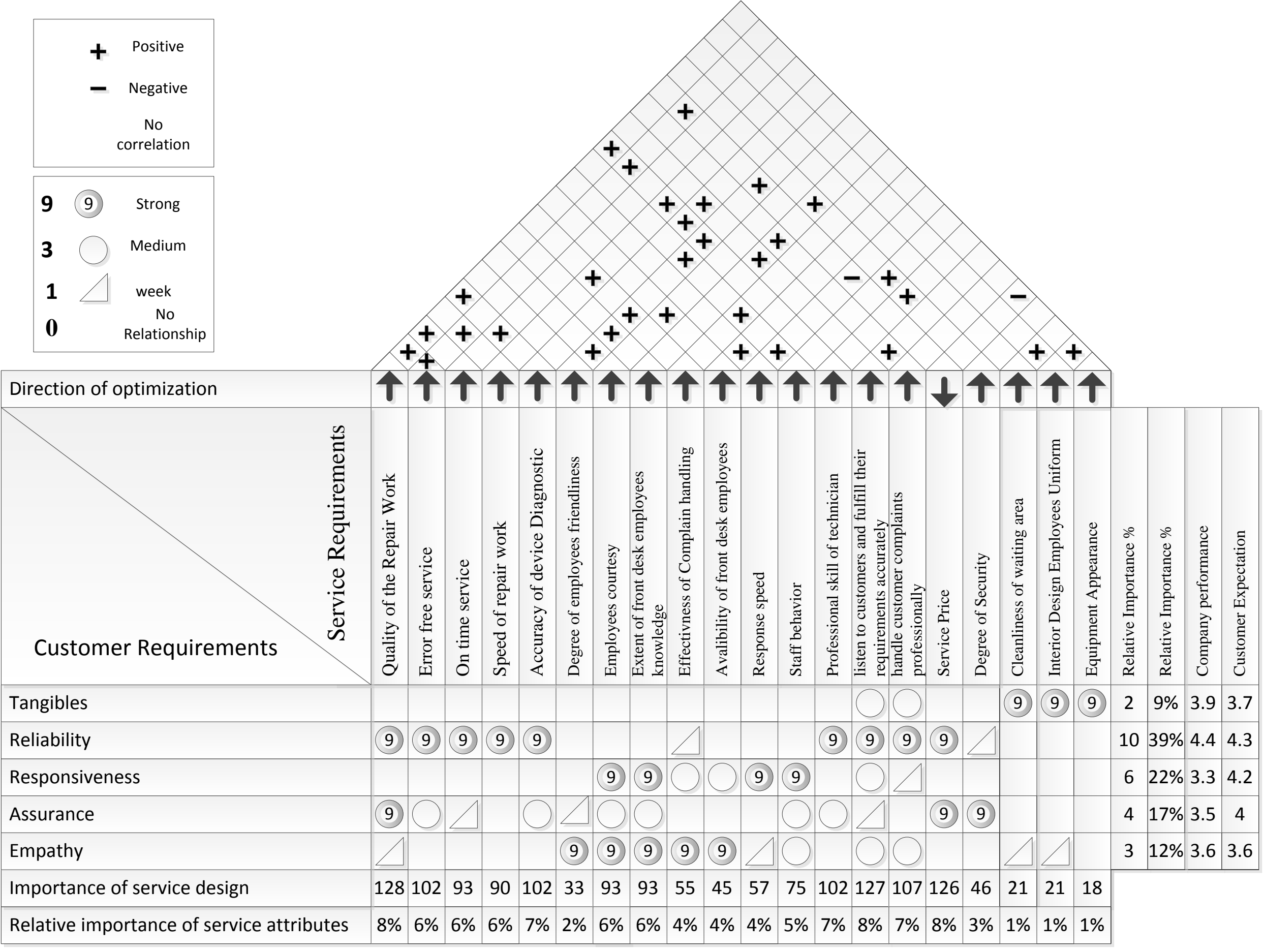

Figure 5.6 House of Quality for XYZ Company

Since XYZ Company aims to improve its performances through increasing the number of satisfied customers, then calculating the importance level is considered the most critical step to achieve the company desired goal. The importance weight reflects the magnitude of customer demand for each technical characteristic. Figure 5.6 illustrates the value for technical attribute. The first attribute "the quality of repair work" has ranked number one, meaning it's the most important characteristic – the quality of repair work has the highest impact on customer satisfaction. On the other hand, equipment appearance received the lowest score, revealing this attribute has the lowest influence on satisfaction level.

5.4.4. Measure perception level:

The customer perception survey has been constructed to capture the customer attitude regarding the provided services at XYZ Company. In the previous chapter, the author highlighted 28 items, which represents the drivers of customer satisfaction in the computer service industry. However, it's difficult to make the customers evaluate all 28 items and some of those requirements do not reflect the actual customer needs of XYZ Company. For these reasons, 17 items have been selected which reflect the customer major requirements. As a final step, the customers have been appealed to rate them on a five-point Likert scale, ranging from 1 (extremely disagree) to 5 (extremely agree). A paper-based survey has been used, meaning the surveys are handed out and completed manually. After that, the data has been saved on Excel spreadsheets and analyzed by using Minitab software.

Table 5.3 illustrates the average perception level for each item. The customers of XYZ Company show a positive attitude regarding item 1 which exhibits accuracy level of delivering service right the first time and item 2 which reflects accuracy level of delivering the service at the promised time. Since the average perception level for both items is over four, the delivered value of those features is considered a good quality from a customer perspective. Similarly, the customers agree that their personal information is safe and the employees respect their privacy. The same is true with the features 10, 11 and 15 – the customer agrees that the operation hours are convenient and the employees are dressed professionally. However, customers think the personal attention of XYZ Company is below average, meaning the front desk employees are poorly trained, which, in return negatively impacts customer satisfaction. In addition, customers feel that the front desk employees don't understand their actual needs which will impact the overall service quality. The perception levels for feature 3, 4, 5, 7, 8, 9, 13 and 16 fall in the neutral zone – customers feel that the delivered quality is neither good nor bad.

Unfortunately, the customer perception survey alone doesn't provide overall customer attitude regarding provided services. It cannot be used as a satisfaction predictor because the satisfaction happens only when the perception level exceeds the satisfaction level. For that reason, the author has deployed the SERVQUAL instrument to find the gap between customer perception and expectation (Gap 5).

Table 5.3 Customer Perception Level

| Item number | Descriptions | Average | Variance |
|---|---|---|---|
| **1** | Accuracy level of delivering service the service right at the first time | 4.444444444 | 0.35628858 |
| **2** | Accuracy level of delivering the service at the promised time | 4.320987654 | 0.5 |
| **3** | Employees attitude toward the customers | 3.209876543 | 0.573302469 |
| **4** | Speed level of response | 3.222222222 | 0.548611111 |
| **5** | Employees availability to assist the customer | 3.567901235 | 0.582561728 |
| **6** | Safety level of customer information | 4.382716049 | 0.484375 |

| Item number | Descriptions | Average | Variance |
|---|---|---|---|
| **7** | Reasonable repair cost | 3.086419753 | 0.520833333 |
| **8** | The level of employees Courtesy | 3.24691358 | 0.722029321 |
| **9** | The level of employees knowledge | 3.469135802 | 0.638888889 |
| **10** | The level of convenience of operating hours | 4.222222222 | 0.430362654 |
| **11** | The level of convenience of service location | 4.24691358 | 0.50617284 |
| **12** | The level personal attention | 2.962962963 | 0.527006173 |
| **13** | The simplicity of the language that is used in communication | 3.827160494 | 0.620177469 |
| **14** | The level of understanding customer needs | 2.790123457 | 0.583333333 |
| **15** | Employees appearances | 4.037037037 | 0.513695988 |
| **16** | Comfort level of waiting room | 3.777777778 | 0.637152778 |
| **17** | Visual aspect of Equipment | 3.864197531 | 0.629436728 |

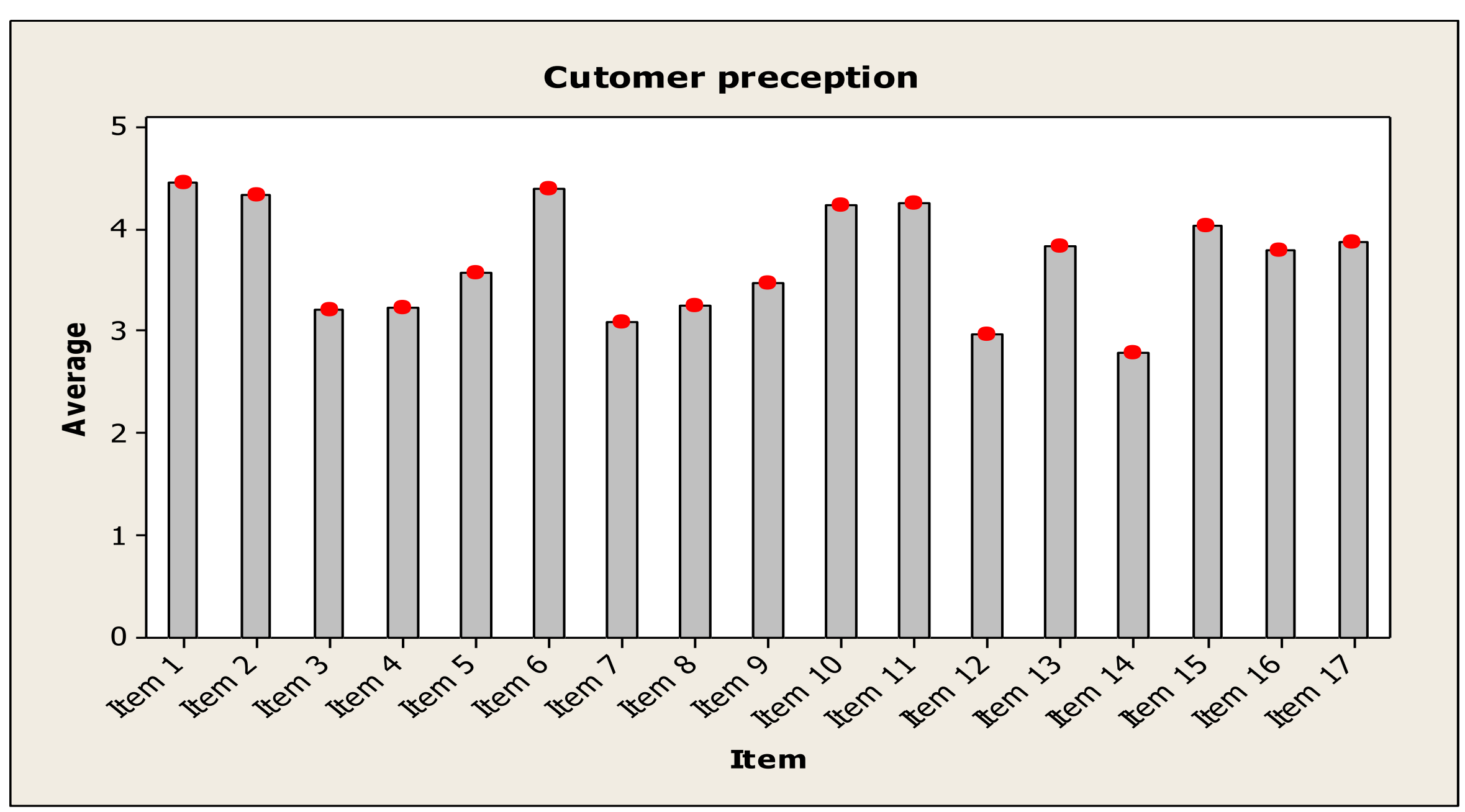

Figure 5.7 Perception Levels for Each Item

5.4.5 Define Measurement Metrics:

In the current study, measurement variables can be categorized into two branches. The first branch encompasses all variable, which are considered critical to customers. Those variables are known as the independent variables. The satisfaction level presents the second type of variables, which is also called dependent variables.

Two fundamental groups of independent variables have been highlighted in this study. The first group is associated with customer expectation which is subdivided into two subgroups. The customer requirement variable is the first subgroup, which consists of 17 variables as illustrated in table 5.4. The letter $E_n$ was used to indicate the expected level for each item, and the letter n was used to denote item number. The importance weight is the second subgroup, which consists of five variables (See table 5.4). Customer perception variable is the second group of independent variables, which is comprised of 17 items as exhibited in table 5.4. The letter P was used to symbolize perception level for each item.

The satisfaction level for each feature can be found by subtraction of the perceived level from the expected level. The satisfaction level for each dimension can be calculated by taking the average score for the items which compromised the dimension. For instance, the satisfaction level for the assurance dimension is found by taking the average score for item 6, 7, 8, and 9. Unfortunately, those dimensions are unequally weighted. That reveals necessity for the second group. Table 5.4 illustrates the dependent variables for this study.

Table 5.4 Dependent Variables

| Dimension | Item number | Description | Perception | Expectation | Satisfaction |
|---|---|---|---|---|---|
| Reliability | 1 | Accuracy level of delivering service the service right at the first time | $P_1$ | $E_1$ | $Y_1 = P_1 - E_1$ |
| | 2 | Accuracy level of delivering the service at the promised time | $P_2$ | $E_2$ | $Y_2 = P_2 – E_2$ |
| Responsiveness | 3 | Employees attitude toward the customers | $P_3$ | $E_3$ | $Y_3 = P_3 – E_3$ |
| | 4 | Speed level of response | $P_4$ | $E_4$ | $Y_4 = P_4 – E_4$ |
| | 5 | Employees availability to assist the customer | $P_5$ | $E_5$ | $Y_5 = P_5 – E_5$ |
| Assurance | 6 | Safety level of customer information | $P_6$ | $E_6$ | $Y_6 = P_6 – E_6$ |
| | 7 | reasonable repair cost | $P_7$ | $E_7$ | $Y_7 = P_7 – E_7$ |
| | 8 | The level of employees Courtesy | $P_8$ | $E_8$ | $Y_8 = P_8 – E_8$ |

| Dimension | Item number | Description | Perception | Expectation | Satisfaction |
|---|---|---|---|---|---|
| | 9 | The level of employees knowledge | $P_9$ | $E_9$ | $Y_9 = P_9 - E_9$ |
| Empathy | 10 | The level of convenience of operating hours | $P_{10}$ | $E_{10}$ | $Y_{10} = P_{10} - E_{10}$ |
| | 11 | The level of convenience of service location | $P_{11}$ | $E_{11}$ | $Y_{11} = P_{11} - E_{11}$ |
| | 12 | The level personal attention | $P_{12}$ | $E_{12}$ | $Y_{12} = P_{12} - E_{12}$ |
| | 13 | he simplicity of the language that is used in communication | $P_{13}$ | $E_{13}$ | $Y_{13} = P_{13} - E_{13}$ |
| | 14 | The level of understanding customer needs | $P_{14}$ | $E_{14}$ | $Y_{14} = P_{14} - E_{14}$ |
| Tangibles | 15 | Employees appearances | $P_{15}$ | $E_{15}$ | $Y_{15} = P_{15} - E_{15}$ |
| | 16 | Comfort level of waiting room | $P_{16}$ | $E_{16}$ | $Y_{16} = P_{16} - E_{16}$ |
| | 17 | Visual aspect of Equipment | $P_{17}$ | $E_{17}$ | $Y_{17} = P_{17} - E_{17}$ |

### 5.4.6. Reliability Analysis:

To ensure the collected data are trustworthy, a reliability analysis test has been conducted to measure the internal consistency of both perception and satisfaction surveys. For that purpose a Cronbach's alpha has been used, which is the most widely used statistical tool for assessing the overall reliability of the entire scale. Cronbach's alpha varies from 0 to 1- the closer to 1, the more reliable survey and vice versa. Larcker and Fornell claim the value of alpha should be more than 0.6 in order to consider the questionnaire reliable.

In this study, the overall reliability scale for customer perception is 0.7242 which exceeds the threshold of 0.6. This indicates that the customer perception survey is reliable. Similarly, the customer perception survey is reliable since the overall reliability is over 0.6. Table 5.5 and Table 5.6 illustrate the reliability of the scales when each item is deleted. Since the values of Cronbach's alpha are over 0.6 for all items, all 17 items are accepted for analysis. In summary, the reliability test indicates both surveys are trustworthy.

| Cronbach's Alpha for Customer Perception Survey | 0.7062 |
| --- | --- |
| Cronbach's Alpha for Customer Expectation Survey | 0.7242 |

Table 5.5 Reliability Test for Customer Expectation

| Variable | Mean | StDev | Total Corr | Corr | Alpha |
|---|---|---|---|---|---|
| **Q1** | **58.235** | **5.114** | **0.2089** | **0.1775** | **0.7011** |
| **Q2** | **58.358** | **5.117** | **0.1445** | **0.1103** | **0.7084** |
| **Q3** | **59.469** | **4.838** | **0.5213** | **0.4687** | **0.6680** |
| **Q4** | **59.457** | **5.003** | **0.3062** | **0.3003** | **0.6921** |
| **Q5** | **59.111** | **4.891** | **0.4059** | **0.3113** | **0.6803** |
| **Q6** | **58.296** | **5.041** | **0.2759** | **0.2716** | **0.6952** |
| **Q7** | **59.593** | **4.896** | **0.4737** | **0.3840** | **0.6747** |
| **Q8** | **59.432** | **4.845** | **0.4379** | **0.4276** | **0.6759** |
| **Q9** | **59.210** | **4.931** | **0.3516** | **0.3261** | **0.6867** |
| **Q10** | **58.457** | **5.055** | **0.2609** | **0.2083** | **0.6966** |
| **Q11** | **58.432** | **5.104** | **0.1653** | **0.2623** | **0.7063** |
| **Q12** | **59.716** | **5.065** | **0.2214** | **0.2538** | **0.7007** |
| **Q13** | **58.852** | **5.025** | **0.2481** | **0.2065** | **0.6984** |
| **Q14** | **59.889** | **4.952** | **0.3588** | **0.2673** | **0.6863** |
| **Q15** | **58.642** | **4.948** | **0.3925** | **0.3347** | **0.6832** |
| **Q16** | **58.901** | **5.178** | **0.0443** | **0.2358** | **0.7201** |
| **Q17** | **58.815** | **5.067** | **0.1849** | **0.2607** | **0.7055** |

Table 5.6 Reliability Test for Customer Perception

| Variable | Mean | StDev | Total Corr | Corr | Alpha |
|---|---|---|---|---|---|
| **Item1** | **62.383** | **4.654** | **0.1339** | **0.2567** | **0.7259** |
| **Item2** | **62.481** | **4.580** | **0.2246** | **0.2048** | **0.7195** |
| **Item3** | **62.704** | **4.597** | **0.1318** | **0.2286** | **0.7314** |
| **Item4** | **62.432** | **4.508** | **0.3537** | **0.2962** | **0.7080** |
| **Item5** | **62.543** | **4.511** | **0.2673** | **0.3241** | **0.7165** |
| **Item6** | **62.593** | **4.494** | **0.3063** | **0.3303** | **0.7122** |
| **Item7** | **62.580** | **4.301** | **0.5463** | **0.3627** | **0.6845** |
| **Item8** | **62.691** | **4.601** | **0.1814** | **0.3515** | **0.7236** |
| **Item9** | **63.012** | **4.551** | **0.2633** | **0.2413** | **0.7161** |
| **item10** | **62.938** | **4.434** | **0.3544** | **0.3024** | **0.7071** |
| **Item11** | **63.346** | **4.523** | **0.3494** | **0.2860** | **0.7087** |
| **Item 12** | **63.296** | **4.267** | **0.6006** | **0.5343** | **0.6780** |
| **Item13** | **63.247** | **4.471** | **0.3550** | **0.2793** | **0.7072** |
| **Item14** | **63.049** | **4.658** | **0.0880** | **0.2148** | **0.7317** |
| **Iem15** | **63.346** | **4.525** | **0.3271** | **0.2353** | **0.7104** |
| **Item16** | **62.605** | **4.438** | **0.4657** | **0.3449** | **0.6973** |
| **Item17** | **63.198** | **4.540** | **0.2878** | **0.2039** | **0.7139** |

### 5.4.7. Use SERVQUAL to measure the satisfaction level:

In previous steps, both perceived and expected levels have been calculated. This step is specified to determine the shortfalls in the service quality in XYZ Company. As discussed previously, SERVQUAL is considered the most effective tool to pinpoint the gap between perception and expectation. For that particular reason, the author has deployed it in this framework to highlight the root causes of customer dissatisfaction.

At the beginning, the level of satisfaction for each item is calculated by subtracting the perceived level from the expected level. The positive values indicate the company achieved its goal by fulfilling customer requirements. The negative values reveal the deficiency of the service quality – the more negative score, the more unsatisfied customers. Column three in Table 5.7 through 5.11 illustrates the gap in delivered service. After that, the satisfaction level for each dimension has been calculated by taking the average gap score for the items that compromises it. Since the impact of those dimensions on customers' judgment is not equal, it is necessary to calculate the weighted dimension score of those dimensions. The last row of the tables below (Table 5.7- Table 5.11) illustrates the weighted satisfaction score for each dimension. To gain proper insight about XYZ Company performance regarding customer requirements, the author will examine each service dimension in the following sections.

#### 5.4.7.1 Reliability:

The reliability dimension reflects the company ability to deliver the service right the first time at the promised time. In this study, the reliability dimension encompasses the first two items. Item1 reflects XYZ Company's ability to deliver the service right at the specified time.

Since the overall customers’ expectations is less than the customers’ perception, the customers are satisfied with this feature. XYZ Company succeeds in delivering error-free service, which involves efficient diagnosing and repairing of hardware and software issues. Likewise, since the gap between customer perception and expectation is positive (0.024), the customers agree that they received the service at the promised time. The average score of reliability dimension is 0.037, meaning customers are satisfied with the quality of this dimension. The previous score doesn’t reflect the actual level of satisfaction with reliability dimension because those dimensions are unequally weighted. The average weighted score for this dimension is 1.47. The table below summarizes the results for reliability dimension.

Table 5.7 Reliability Dimension- Result Summary

| **Reliability** | | | | |
|---|---|---|---|---|
| **Item number** | Customer expectation | customer perception | Gap between customer expected and perceived value | Average importance score |
| **Item 1** | 4.395061728 | 4.444444444 | 0.049382716 | 39.69512195 |
| **Item 2** | 4.296296296 | 4.320987654 | 0.024691358 | |
| **Average unweight Reliability score** | | | 0.037037037 | |
| **Average weighted Reliability score** | | | 1.470189702 | |

5.4.7.2. Responsiveness

Responsiveness ranks the second most important dimension from a customer perspective. Zeithamal, Parasuraman, and Berry defined this dimension as the company willingness to serve their customers and provide them with promoted services. In this current study, the responsiveness dimension consists of items 3 through 5. The employees' attitude toward the customers is represented by item 3. The average gap score for item 3 is -0.864 which indicates that XYZ Company performs poorly in this feature. Either front desk employees or technicians didn't show a positive attitude towards the current customers. The case is even worse for item 4 since the gap between customer perception and expectation is higher (-1.12). The previous result shows the customers are not satisfied with employees' response speed. Employees' availability to help the customer is the last item in this dimension. In addition, XYZ Company does not fulfill the customer expectation for this item since the gap is negative value (-0.67). The overall average gap for the responsiveness dimension is -0.88 which indicates that the customers are not satisfied with the provided quality. Table 5.8 summarizes the results for this dimension.

Table 5.8 Responsiveness Dimension- Result Summary

| Responsiveness | | | | |
|---|---|---|---|---|
| **Item number** | Customer expectation | customer perception | Gap between customer expected and perceived value | Average importance score |
| **Item 3** | 4.074074074 | 3.209876543 | -0.864197531 | 22.19512195 |
| **Item 4** | 4.345679012 | 3.222222222 | -1.12345679 | |
| **Item 5** | 4.234567901 | 3.567901235 | -0.666666667 | |
| **Average un-weighted Responsiveness score** | | | -0.884773663 | |
| **Average weighted Responsiveness score** | | | -19.63765934 | |

5.4.7.3. Assurances:

As described previously, assurance dimension represents employees' competence to gain customer trust and confidence. In this study, this dimension consists of four items representing four different areas: courtesy, credibility, competence and security. Often, the personal computers contain sensitive information, so the misuse of that information will cause serious harm to the customers and extreme dissatisfaction with the company in return. The safety level of customer information is represented by item 6. For this item, the perceived level exceeds the expected level that indicates the customers agree that their personal information is safe and the employees are trustworthy.

Based on table 5.9, the average gap score for item 7 is -1.11, which means the customers are extremely unsatisfied with service cost – the service cost is unreasonable. Similarly, since the gap between customer perception and customer expectation is negative (-0.84), the customers think that the employees of XYZ Company are not courteous enough. Item 9 represents the employees' knowledge that involves technical and social skills. The customers think the employees' knowledge is not what they expect. From table 5.9, the average score for this dimension is -0.51, meaning customers are not satisfied with the provided service for this dimension.

Table 5.9 Assurances Dimension- Result Summary

| **Assurances** | | | | |
|---|---|---|---|---|
| **Item number** | Customer expectation | customer perception | Gap between customer expected and perceived value | Average importance score |
| **Item 6** | 4.185185185 | 4.382716049 | 0.197530864 | 16.82926829 |
| **Item 7** | 4.197530864 | 3.086419753 | -1.111111111 | |
| **Item 8** | 4.086419753 | 3.24691358 | -0.839506173 | |
| **Item 9** | 3.765432099 | 3.469135802 | -0.296296296 | |
| **Average un-weighted Assurances score** | | | -0.512345679 | |
| **Average weighted Assurances score** | | | -8.622402891 | |

5.4.7.4. Empathy:

Empathy dimension represents the company attitude to take care of their customers and provide them with individual attention. In the present study, this dimension consists of five elements. The first element is represented by item 10 which refers to the convenience of operating hours. Since the perceived level is higher than the expected level, the customers are satisfied with the company operating hours. Likewise, the customers are satisfied with item 11 which refers to service location. However, they do not agree that the employees provide them with special care since the gap for item 12 has a negative value. Item13 is the fourth element in this dimension, referring to the simplicity of the language used in communication with customers. Illustrated in table 5.10, the perceived level is higher than the expected level which means the company fulfilled customer demand for this item. On the contrary, customers are dissatisfied with the last item, as they think the company did not understand their needs. The average gap score for this dimension is 0.007, indicating customers are satisfied with the quality of provided services.

Table 5.10 Empathy Dimension- Result Summary

| **Empathy** | | | | |
|---|---|---|---|---|
| **Item number** | Customer expectation | customer perception | Gap between customer expected and perceived value | Average importance score |
| **Item 10** | 3.839506173 | 4.222222222 | 0.382716049 | 12.56097561 |
| **Item 11** | 3.432098765 | 4.24691358 | 0.814814815 | |
| **Item 12** | 3.481481481 | 2.962962963 | -0.518518519 | |
| **Item 13** | 3.530864198 | 3.827160494 | 0.296296296 | |
| **Item 14** | 3.728395062 | 2.790123457 | -0.938271605 | |
| **Average unweighted Empathy score** | | | 0.007407407 | |
| **Average weighted Empathy score** | | | 0.093044264 | |

5.4.7.5. Tangibles:

Tangibles dimension refers to all physical aspects of the company which include physical facility, equipment and employees' appearance. In this study, this dimension covers items 15 through 17. Starting with item 15, the customers agree that employees are dressed professionally. Also, the customers agree that the equipment is visually appealing. However, they think the waiting room is not comfortable since the customer perception is less than the customer expectation. The average score for tangibles dimension is 0.16, indicating the customers are satisfied with the quality of this dimension. The results are summarized in Table 5.11.

Table 5.11 Tangibles Dimension- Result Summary

| **Tangibles** | | | | |
|---|---|---|---|---|
| **Item number** | Customer expectation | customer perception | Gap between customer expected and perceived value | Average importance score |
| **Item 15** | 3.432098765 | 4.037037037 | 0.604938272 | 8.780487805 |
| **Item 16** | 4.172839506 | 3.777777778 | -0.395061728 | |
| **Item 17** | 3.580246914 | 3.864197531 | 0.283950617 | |
| **Average un-weighted Tangibles score** | | | 0.164609053 | |
| **Average weighted Tangibles score** | | | 1.445347787 | |

5.4.8 Analyze the root cause of customer dissatisfaction:

This step is specified to highlight the root causes of customer dissatisfaction in XYZ Company. The current phase relies heavily on the results from the previous steps. The cause and effect diagram has been employed for that particular reason, and five major causes of customer dissatisfaction have been identified at XYZ Company.

The physical environment represents the first root for customer dissatisfaction. As previously illustrated, the customers think the comfort level of the waiting area is not as expected. This type of dissatisfaction can be traced back to two causes: for one reason, the waiting area does not have enough entertainment such as Wi-Fi, TVs, or comfortable chairs. Also, the level of cleanliness is not as expected. The company's physical environment is considered a minor factor of customer dissatisfaction.

Service cost is another cause for dissatisfaction for the XYZ Company. Service cost is considered a superior cause for customer dissatisfaction since it impacts company credibility and hurts customer loyalty in return. Many of the computer service companies underestimate the impact of front desk employees on the overall satisfaction and many of them think providing the customer with efficient diagnostic and repair is more than enough; after all, this is not the truth.

This study highlights an important concept: providing customers with reliable service is not enough if not accompanied with proper customer care from front desk employees. The next three groups of customer dissatisfaction causes are connected to front desk employees. The third root of customer dissatisfaction is linked to employees' response. In many cases, the companies hire inadequate number of employees to reduce the expenses; however, this negatively impacts the customer satisfaction in three different areas. For one, employees will have less time to assist the customers – this will impact employees' courtesy and response speed. Hiring inefficient employees will have similar impact on customer satisfaction. Similarly, XYZ Company hired unqualified employees who do not respond efficiently to customers that cause three issues as illustrated in the figure below.

The front desk employees do not have certain social skills that are considered the fourth cause of customer dissatisfaction. The last cause is connected to staff technical skills that represent the knowledge and competence of engineers, technicians and front desk employees. As illustrated in the previous section, the customers are satisfied with the reliability of the provided service that indicates both the engineers and technicians are efficient; however, many of the customers do not agree that the front desk employees are knowledgeable and able to understand their needs. Often, the only person who interacts with a customer is the front desk employee – that should indicate the necessity to train him/her to convey the customers' trust and confidence.

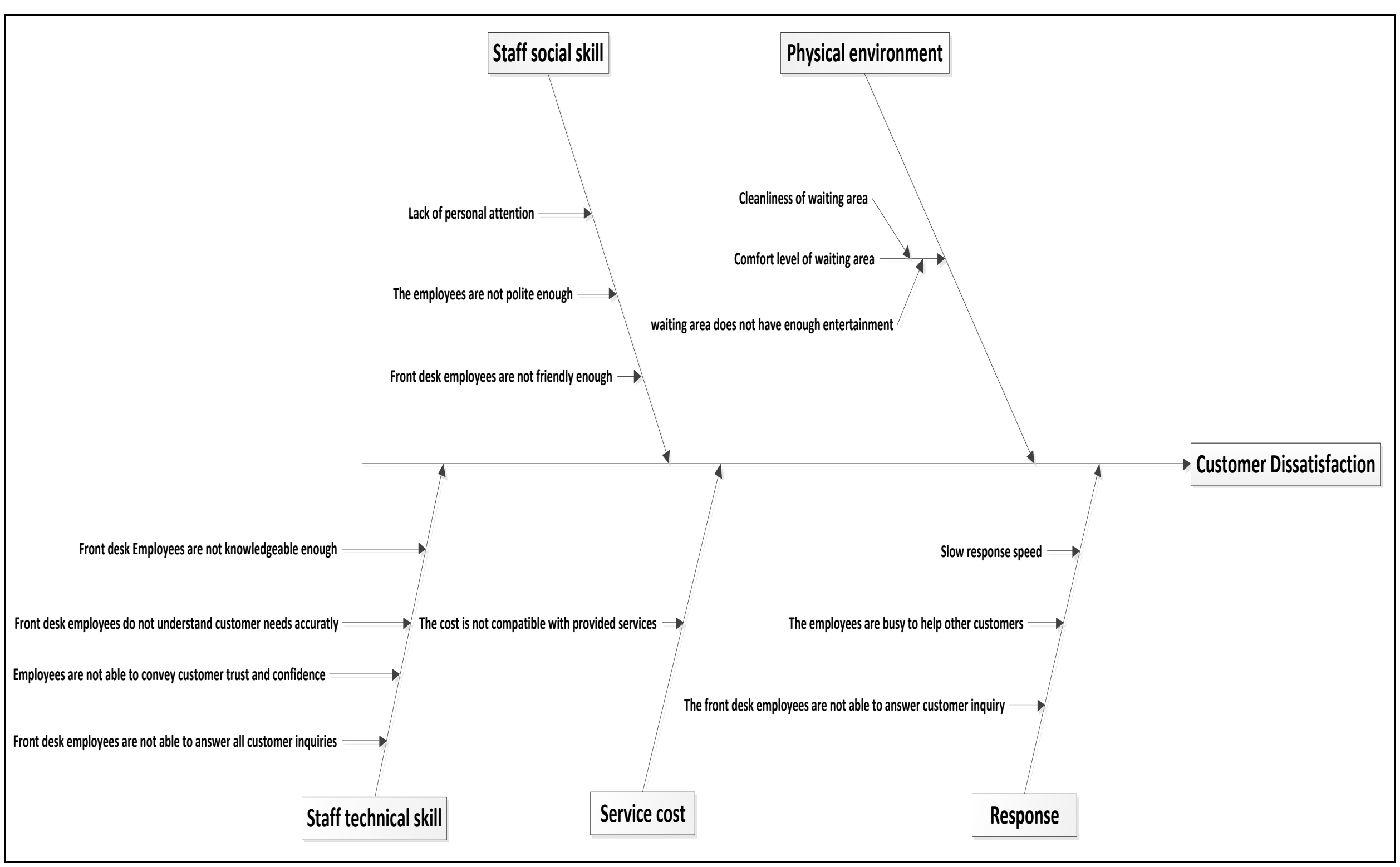

Figure 5.8 Cause and Effect Diagram for XYZ Company

In conclusion, the fundamental reason for customer dissatisfaction at XYZ Company is front desk employees since the customers never complain about the reliability of provided services. The company technicians are able to diagnose and repair hardware and software issues efficiently and deliver the service correctly the first time, indicating XYZ Company hires highly qualified technicians and computer engineers. However, it is not enough to achieve customer loyalty since the front desk employees do not have certain social and technical skills to win customer trust, handle complaints, or even respond to complaints correctly. As described previously, now-days the customers become more demanding and require excellent customer service – providing them with reliable service without proper customer care is not enough to maintain customer loyalty. In order for the company to survive among competitive opponents, XYZ Company needs to hire competent and highly skilled front desk employees.

# CHAPTER SIX:
# CONCLUSION

## 6.1 Introduction

As previously described, the computer service industry sector aggressively suffers from customer decay, yet many computer service companies find it is easier to attract new customers rather than keeping current customers, which is opposite to the truth.

As the author illustrates in literature review, marketing studies have proved that keeping current customers is four to seven times easier than attracting new customers. If this is true, then why do many computer service companies think the opposite? An enormous amount of literature has been reviewed and extensive investigation has been conducted by the author to answer this vital question. Since most of computer maintenance product characteristics are intangible, then most of traditional quality tools cannot be used. Measuring the quality of delivered service provides the company with the insights to delight and retain customers. By determining customers' needs and fulfilling them, the computer service companies not only retain satisfied customers, but also reduce the cost of accruing new customers. In the other words, satisfying the customer is considered vital for company long-term success.

Many studies show the reason for companies' failure to sustain customer satisfaction is a lack of a robust model to measure customer satisfaction level (Vivra, 1997). In general, the services are heterogeneous, inseparable and intangible, which explains why a number of companies find it difficult to measure quality of the service and as a result, fail to satisfy their customer requirements.

## 6.2 Answering Research Questions

This study has been conducted to answer two groups of questions. The first group is comprised of the first two questions of this study. The first part of chapter four has specified to answer question one. Twenty eight items have been highlighted which represent the key quality characteristics at computer service industry. Those items have been grouped into five distinct dimensions, which are: Reliability, Tangibles, Assurance, Responsiveness and Empathy. The second part of chapter four is intended to answer question two, which involves developing a robust framework to measure customer satisfaction at computer service industry. The second group is represented by question three through five. They reflect the second part of study that involves deploying the proposed framework in XYZ Company. Section 6.4 will specifically answer questions three through five.

## 6.3 Framework development

The author has developed a framework to measure and analyze customer satisfaction at computer service industries, since no study has been conducted for that particular purpose. Many quality tools have been deployed in this framework along with SERVQUAL. In the first step, SIPOC diagram has been employed, along with a flow chart to gain full understanding about how the customer interacts with the service provider and what the process key players are.

Voice of customer has been used for two specific reasons, which identify customer requirements first and measure expectation levels for each service characteristic second. The customer requirements should translate into technical requirements – for that purpose house of quality has been employed which is considered the most efficient tool. Customer perception surveys have been constructed to measure delivered value for each feature “perceived level.”

Measuring the reliability of both customer perception and customer satisfaction is considered a vital step since those surveys have sort variability. Cronbach’s Alpha is the most effective statistical tool to test the reliability. SERVQUAL instrument has been deployed to measure satisfaction level. At the end, cause and effect diagram has adapted to highlight the root causes of customer dissatisfaction. To check the effectiveness of this frame work, it has been deployed to measure customer satisfaction and highlight root causes of dissatisfaction. In summary, this study provides the computer service with a framework not only to measure and analyze customer satisfaction but also to translate customer requirement into technical plan.

## 6.4 Result summary for XYZ Company

As previously stated, XYZ Company is one of those companies suffering from customer decays due to lacking a robust measurement system to assess the customer satisfaction. Deploying current framework in XYZ Company will provide the whole picture about what the customers are expected and what the major causes of customer dissatisfaction are. To measure the level of expectation, the customers have been solicited to evaluate the quality of provided service. Seventeen questionnaires have been constructed to measure the level of expectation in five sectors. The reliability of provided service has the highest expected level (4.34) followed by responsiveness (4.21) and assurance (4.05).

The physical aspect has the lowest expected level that means satisfying this dimension is easier than other. Those dimensions vary radically when it comes to importance weight.

According to the satisfaction surveys, the reliability dimension is considered the most importance score (39.6) among the five dimensions which means this dimension has the highest impact on customer satisfaction as well as it is harder to satisfy than the others. With importance weight 22.1, the responsiveness has the second most important dimension followed by assurance (16.8) and empathy (12). Similarly, tangible dimension has ranked the least important dimension which means it has the least impact on overall satisfaction.

To measure the satisfaction level, the customers have been asked again to evaluate the perceived service for the same 17 features. The satisfaction level (Y) represents the difference between perceived and delivered service. The overall customer satisfaction with provided service is -25 which indicates customer dissatisfaction with provided service. According to the last step of the framework, the root causes are connected to five major sectors which are: staff social skills, staff technical skills, staff response, physical environment and service cost. The case and effect diagram illustrates that most of the causes are linked to front desk employees. Even though the customer expectations for reliability dimension are high, XYZ Company satisfies their customers. That means XYZ Company realizes the criticalness of this dimension by hiring qualified technicians that are able to deliver the service right the first time. However, the company administrators neglect the importance of the front desk employees which results in dissatisfaction with the provided service. This reveals a fundamental concept which is providing the customer with a reliable service does not guarantee customer satisfaction. To increase customer satisfaction, XYZ should design a training program to improve courtesy, social skills, response and technical knowledge of front desk employees. Adding some excitement to the waiting area will have a positive influence on satisfaction level.

Finally, reducing the cost will have a huge impact on customer satisfaction since it is considered a one-dimension requirement.

# APPENDIX A:
# IRB APPROVAL LETTER

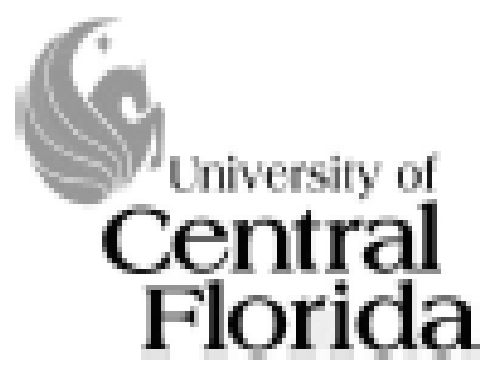

University of Central Florida Institutional Review Board
Office of Research & Commercialization
12201 Research Parkway, Suite 501
Orlando, Florida 32826-3246
Telephone: 407-823-2901 or 407-882-2276
www.research.ucf.edu/compliance/irb.html

## Approval of Exempt Human Research

From: **UCF Institutional Review Board #1**
**FWA00000351, IRB00001138**

To: **Mohammed Abboodi**

Date: **October 10, 2013**

Dear Researcher:

On 10/10/2013, the IRB approved the following activity as human participant research that is exempt from regulation:

| | |
|---|---|
| Type of Review: | Exempt Determination |
| Project Title: | Framework to measure and analyze customer satisfaction at computer service industry Using Lean Six Sigma |
| Investigator: | Mohammed Abboodi |
| IRB Number: | SBE-13-09678 |
| Funding Agency: | |
| Grant Title: | |
| Research ID: | N/A |

This determination applies only to the activities described in the IRB submission and does not apply should any changes be made. If changes are made and there are questions about whether these changes affect the exempt status of the human research, please contact the IRB. When you have completed your research, please submit a Study Closure request in iRIS so that IRB records will be accurate.

In the conduct of this research, you are responsible to follow the requirements of the Investigator Manual.

On behalf of Sophia Dziegielewski, Ph.D., L.C.S.W., UCF IRB Chair, this letter is signed by:

Signature applied by Joanne Muratori on 10/10/2013 11:22:07 AM EDT

Joanne Muratori

IRB Coordinator

# APPENDIX B:
# CUSTOMER EXPECTATION SURVEY

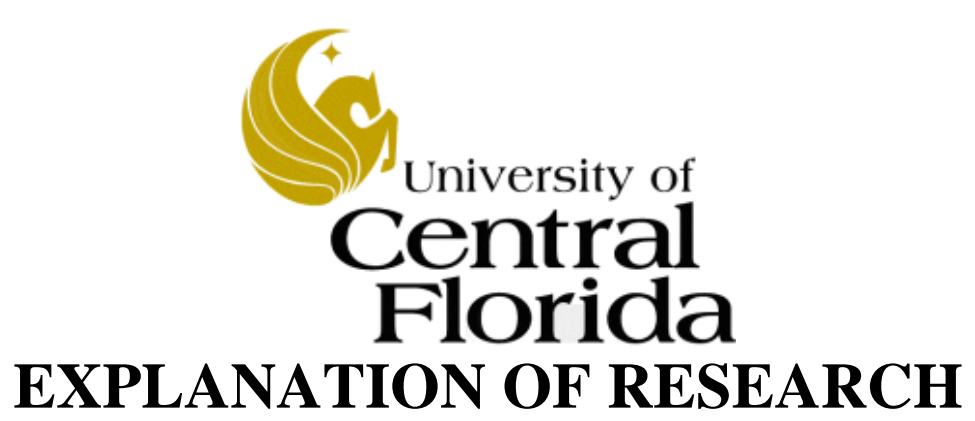

**EXPLANATION OF RESEARCH**

Title of Project: Framework to Measure and Analyze Customer Satisfaction at Computer Services Industry Using Lean Six Sigma.

Principal Investigator: Mohammed Abboodi

Other Investigators: N/A

Faculty Supervisor: Dr. Ahmad Elshennawy

You are being invited to take part in a research study. Whether you take part is up to you.

- The main purpose of this study is to develop framework to measure and analyze customer satisfaction in computer services industry. This framework will assist the company administrators to improve their company performances through increasing their customer satisfaction.
- You will be asked to complete short surveys regarding either received or expected service at the Computer Repair Company while you are waiting to receive service.
- The time needed to complete the survey questions will be about 6 minutes

You must be 18 years of age or older to take part in this research study.

**Study contact for questions about the study or to report a problem:** If you have questions, concerns, or complaints, please contact *Mohammed Abboodi, Graduate Student,Industerial engineering , College o f Engineering and computer scince, (407) 309-0552 or by email Abboodi@knights.ucf.edu.*

**IRB contact about your rights in the study or to report a complaint:** Research at the University of Central Florida involving human participants is carried out under the oversight of the Institutional Review Board (UCF IRB). This research has been reviewed and approved by the IRB. For information about the rights of people who take part in research, please contact: Institutional Review Board, University of Central Florida, Office of Research & Commercialization, 12201 Research Parkway, Suite 501, Orlando, FL 32826-3246 or by telephone at (407) 823-2901.

# Computer services customer Expectation Survey

1. We request your help. Please complete the following 17 descriptive statements relate to your expectations about the features should be possed by computer services company. Thank you for your time.

| | | (1 strongly disagree, 5 strongly agree) | | | | |
|---|---|---|---|---|---|---|
| | | 1 | **2** | **3** | **4** | **5** |
| **Reliability** | 1. Excellent computer Services Company will deliver the services right at the first time. | ☐ | ☐ | ☐ | ☐ | ☐ |
| | 2. At Excellent computer Services Company the customers will receive the service at the promised time. | ☐ | ☐ | ☐ | ☐ | ☐ |
| | | | | | | |
| **Responsiveness** | 3. The employees of excellent computer service company will always be willing to help the customer | ☐ | ☐ | ☐ | ☐ | ☐ |
| | 4. The employees of excellent computer Services Company response to the customers quickly. | ☐ | ☐ | ☐ | ☐ | ☐ |
| | 5. The employees of excellent computer Services Company will be available to help the customers. | ☐ | ☐ | ☐ | ☐ | ☐ |
| | | | | | | |
| **Assura** | 6. The customers feel their personal information is secure at excellent computer Services Company. | ☐ | ☐ | ☐ | ☐ | ☐ |

| | | (1 strongly disagree, 5 strongly agree) | | | | |
|---|---|---|---|---|---|---|
| | 7. At Excellent computer Services Company, the cost of repair is reasonable | ☐ | ☐ | ☐ | ☐ | ☐ |
| | 8. The employees of excellent computer Services Company will be courteous. | ☐ | ☐ | ☐ | ☐ | ☐ |
| | 9. The employees of excellent computer Services Company have the knowledge to answer customer questions. | | | | | |
| | | | | | | |
| Empathy | 10. The business hours will be convenient at excellent computer Services Company. | ☐ | ☐ | ☐ | ☐ | ☐ |
| | 11. The parking lot will be near to the service location at excellent computer Services Company. | ☐ | ☐ | ☐ | ☐ | ☐ |
| | 12. The employees will give you the expected attention at excellent computer Services Company. | ☐ | ☐ | ☐ | ☐ | ☐ |
| | 13. The employees of excellent computer Services Company will use a simple language while they communicated with their customers. | ☐ | ☐ | ☐ | ☐ | ☐ |
| | 14. The employees of excellent computer Services Company will understand the their customers' specific needs | ☐ | ☐ | ☐ | ☐ | ☐ |
| | | | | | | |
| Tangib | 15. The employees of excellent computer Services Company will dress professionally | ☐ | ☐ | ☐ | ☐ | ☐ |

| | | (1 strongly disagree, 5 strongly agree) | | | | |
|---|---|---|---|---|---|---|
| | 16. Waiting area of excellent computer Services Company will be comfortable | ☐ | ☐ | ☐ | ☐ | ☐ |
| | 17. The excellent computer Services Company will have a modern looking equipment | ☐ | ☐ | ☐ | ☐ | ☐ |

2. It is important to us to know how the following features are important to you, so please answer these questions: The total available point is 100, so please make sure that the allocated points add up to 100. For example, if you give the feature 18 -21 the following score (20, 30, 10, 10), the available points for the feature (22) is 20.

| **Feature** | Score (0 to 100) |
|---|---|
| | |
| 18. How important the appearance of the company, personal and equipment is to you. | |
| 19. How important the computer services company ability to provide the promised service accurately is to you. | |
| 20. How important the service responsiveness is to you. | |
| 21. How important the service assurance is to you. | |
| 22. How important the service empathy is to you. | |

Thank you very much for taking the time to complete this survey. Your feedback is valued and very much appreciated!

# APPENDIX C:
# CUSTOMER PERCEPTIONS SURVEY

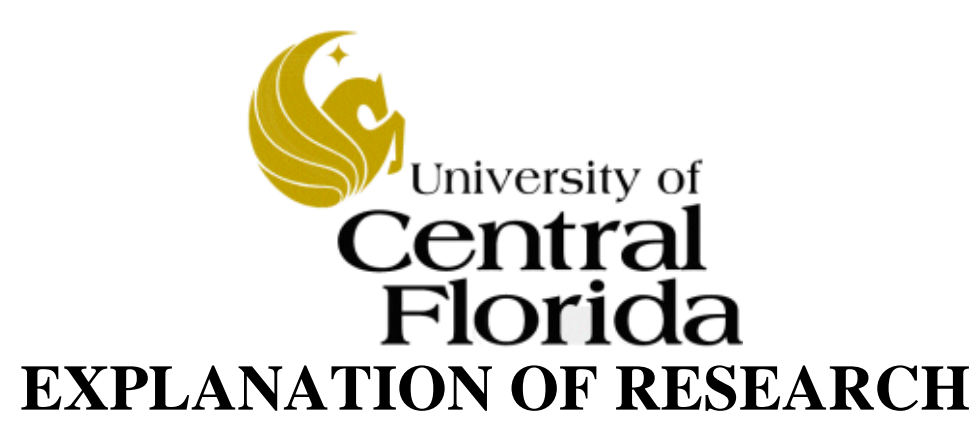

**EXPLANATION OF RESEARCH**

Title of Project: Framework to Measure and Analyze Customer Satisfaction at Computer Services Industry Using Lean Six Sigma.

Principal Investigator: Mohammed Abboodi

Other Investigators: N/A

Faculty Supervisor: Dr. Ahmad Elshennawy

You are being invited to take part in a research study. Whether you take part is up to you.

- The main purpose of this study is to develop framework to measure and analyze customer satisfaction in computer services industry. This framework will assist the company administrators to improve their company performances through increasing their customer satisfaction.
- You will be asked to complete short surveys regarding either received or expected service at the Computer Repair Company while you are waiting to receive service.
- The time needed to complete the survey questions will be about 6 minutes

You must be 18 years of age or older to take part in this research study.

**Study contact for questions about the study or to report a problem:** If you have questions, concerns, or complaints, please contact *Mohammed Abboodi, Graduate Student,Industerial engineering , College o f Engineering and computer scince, (407) 309-0552 or by email Abboodi@knights.ucf.edu.*

**IRB contact about your rights in the study or to report a complaint:** Research at the University of Central Florida involving human participants is carried out under the oversight of the Institutional Review Board (UCF IRB). This research has been reviewed and approved by the IRB. For information about the rights of people who take part in research, please contact: Institutional Review Board, University of Central Florida, Office of Research & Commercialization, 12201 Research Parkway, Suite 501, Orlando, FL 32826-3246 or by telephone at (407) 823-2901.

# Computer services customer perception Survey

We request your help. Please complete the following 17 descriptive statements relate to your feeling about our service. Thank you for your time.

| | | (1 strongly disagree, 5 strongly agree) | | | | |
|---|---|---|---|---|---|---|
| | | 1 | 2 | 3 | 4 | 5 |
| Reliability | 1. Service was delivered right at the first time. | ☐ | ☐ | ☐ | ☐ | ☐ |
| | 2. Did you receive the service right at the promised time? | ☐ | ☐ | ☐ | ☐ | ☐ |
| | | | | | | |
| Responsiveness | 3. The employees always be willing to help you | ☐ | ☐ | ☐ | ☐ | ☐ |
| | 4. Did the employees response to your request quickly? | ☐ | ☐ | ☐ | ☐ | ☐ |
| | 5. The employees were available to help you | ☐ | ☐ | ☐ | ☐ | ☐ |
| | | | | | | |
| Assurance | 6. You feel your personal information was secure. | ☐ | ☐ | ☐ | ☐ | ☐ |
| | 7. The cost of repair was reasonable | ☐ | ☐ | ☐ | ☐ | ☐ |
| | 8. The employees were courteous | ☐ | ☐ | ☐ | ☐ | ☐ |
| | 9. The employees had the knowledge to answer your questions | ☐ | ☐ | ☐ | ☐ | ☐ |
| | | | | | | |
| **Empathy** | 10. The business hours were convenient | ☐ | ☐ | ☐ | ☐ | ☐ |
| | 11. The parking lot was near to the service location. | ☐ | ☐ | ☐ | ☐ | ☐ |
| | 12. The employees gave you the expected attention. | ☐ | ☐ | ☐ | ☐ | ☐ |
| | 13. The employees used a simple language while they communicated with you | ☐ | ☐ | ☐ | ☐ | ☐ |
| | 14. The employees understood the your specific needs | ☐ | ☐ | ☐ | ☐ | ☐ |
| | | | | | | |
| **Tangibles** | 15. The employees dressed professionally | ☐ | ☐ | ☐ | ☐ | ☐ |
| | 16. Waiting area was comfortable | ☐ | ☐ | ☐ | ☐ | ☐ |
| | 17. The company has a modern looking equipment | ☐ | ☐ | ☐ | ☐ | ☐ |

2. This part of survey is optional; however it is important for us to understand your overall satisfaction with the services so please answer these questions:

| | | | | | |
|---|---|---|---|---|---|
| 18. Over all, are you satisfy with the quality of the services that provided by our organization? | ☐ | ☐ | ☐ | ☐ | ☐ |
| 19. Would you recommend our services to others? | ☐ | ☐ | ☐ | ☐ | ☐ |
| 20. Would you use our service again | ☐ | ☐ | ☐ | ☐ | ☐ |

Thank you very much for taking the time to complete this survey. Your feedback is valued and very much appreciated!